\pdfoutput=1
\documentclass[11pt,aps,a4paper,eqsecnum,amsmath,amssymb
              ,notitlepage,nofootinbib
              ,floatfix
              ,longbibliography
              ]{revtex4-1}

\usepackage{amsthm}
\usepackage{amsmath}
\usepackage{amsfonts}

\usepackage{tikz}
\usepackage{pgfplots}
\usepgfplotslibrary{patchplots}
\usetikzlibrary{decorations.markings}
\usetikzlibrary{arrows}

\usepackage{braket}
\usepackage{tabularx}
\usepackage{longtable}
\usepackage{booktabs}
\usepackage{graphicx}
\usepackage{color}

\def\tp{\otimes}
\def\Z{\mathbb{Z}}

\def\R{\mathbb{R}}
\def\C{\mathbb{C}}
\def\Wop[#1][#2][#3][#4][#5]{W\left(\left.\begin{array}{cc} #1&#2\\#3&#4\end{array}\right|#5\right)}
\tikzset{middlearrow/.style={decoration={markings, mark= at position 0.6 with {\arrow{#1}},},postaction={decorate}}}
\def\Afun[#1][#2]{A\left(#1; #2\right)}
\def\Afund[#1][#2]{A'\left(#1; #2\right)}
\newcolumntype{b}{>{\centering}X}
\newcolumntype{r}{>{\hsize=.3\hsize}b}
\newcolumntype{s}{>{\hsize=.2\hsize}b}
\newcolumntype{t}{>{\hsize=.1\hsize}b}

\begin{document}
\preprint{}

\title{$\mathbf{Z}_{n}$ clock models and chains of $so(n)_2$ non-Abelian anyons:\\
  symmetries, integrable points and low energy properties}
\author{Peter E. Finch}
\author{Michael Flohr}
\author{Holger Frahm}
\affiliation{%
Institut f\"ur Theoretische Physik, Leibniz Universit\"at Hannover,
Appelstra\ss{}e 2, 30167 Hannover, Germany}


\begin{abstract}
We study two families of quantum models which have been used previously to investigate the effect of topological symmetries in one-dimensional correlated matter.  Various striking similarities are observed between certain $\mathbf{Z}_n$ quantum clock models, spin chains generalizing the Ising model, and chains of non-Abelian anyons constructed from the $so(n)_2$ fusion category for odd $n$, both subject to periodic boundary conditions. In spite of the differences between these two types of quantum chains, e.g.\ their Hilbert spaces being spanned by tensor products of local spin states or fusion paths of anyons, the symmetries of the lattice models are shown to be closely related.  Furthermore, under a suitable mapping between the parameters describing the interaction between spins and anyons the respective Hamiltonians share part of their energy spectrum (although their degeneracies may differ).
This spin-anyon correspondence can be extended by fine-tuning of the coupling constants leading to exactly solvable models.  We show that the algebraic structures underlying the integrability of the clock models and the anyon chain are the same.  For $n=3,5,7$ we perform an extensive finite size study -- both numerical and based on the exact solution -- of these models to map out their ground state phase diagram and to identify the effective field theories describing their low energy behaviour.  We observe that the continuum limit at the integrable points can be described by rational conformal field theories with extended symmetry algebras which can be related to the discrete ones of the lattice models.
\end{abstract}

\maketitle

\newpage
\section{Introduction}
%
Non-Abelian anyons, appearing as fractionalized quasi-particle excitations with exotic statistics appearing in topologically ordered phases of correlated quantum many-body systems such as fractional quantum Hall states or $p+ip$ superconductors \cite{MoRe91,ReRe99,ReGr00}, have attracted tremendous interest in recent years -- not least due to their potential use as resources in quantum computing \cite{Kita03,NSSF08}.  For several one-dimensional systems forming related topological phases these objects have been shown to be realized as topologically protected zero-energy modes localized at boundaries or defects \cite{Kitaev01,AsNa12,Tsve14a,BoFr17}.  In the simplest case of the quantum Ising chain these modes are Majorana anyons, signatures for these have been found in experiments on heterostructures such as semiconductor quantum wires in proximity to superconductors \cite{MZFP12,Deng.etal12}.

In the search for anyons beyond these Majorana zero modes there has been a revival of the interest in $n$-state generalizations of the Ising model, i.e.\ the one-dimensional $\mathbf{Z}_n$ symmetric quantum clock models recently.  For \emph{open} boundary conditions some of these systems have been found to have an $n$-fold degenerate ground state for chains of arbitrary length \cite{Fendley12,Fendley14,ARFGB16}. This non-trivial degeneracy cannot resolved by the action of local symmetry preserving operators and is closely connected to the existence of localized zero energy modes with fractionalized charge.  
\emph{Periodic} boundary conditions (or, more generally, the addition of a coupling between the edges to the open chain Hamiltonian) destroy the ground state degeneracy, but still allow for the study of the low energy behaviour of the correlated system to identify the phases realized as the interaction varied.
Furthermore, some of these models are exactly solvable at certain values of the coupling constants which allows to provide insights beyond what is possible based on the numerical investigation of finite chains.

An alternative approach to study some of the peculiar topological properties of non-Abelian anyons is based on certain deformations of quantum spin chains. Mathematically, the anyons in these models are objects in a braided tensor category equipped with operations describing their fusion and braiding \cite{Kita06}.   Here the fusion rules underly the construction of the many-anyon Hilbert space and determine the local interactions allowed in the lattice model in terms of so-called $F$-moves.  In numerous studies of anyons satisfying the fusion rules of e.g.\ $su(2)_k$, $D(D_3)$, and $so(5)_2$ subject to different short-ranged interactions and in various geometries a variety of critical phases together with the conformal field theories (CFTs) providing the effective description of their low energy properties in gapless phases have been identified \cite{FTLT07,TAFH08,GTKL09,FiFr13,GATH13,Finch.etal14,FiFF14,BrFF16,VeJS17}.  Exactly solvable anyon chains of this type are closely related to integrable two-dimensional classical lattice systems with `interactions round the face' (IRF): the corresponding anyonic Hamiltonian is a member of a family of commuting operators, generated the transfer matrix of the classical model. For example Fibonacci or, more generally, $su(2)_k$ anyon chains with nearest neighbour interactions \cite{FTLT07} can be derived from the critical restricted solid on solid (RSOS) models \cite{AnBF84}. 

Interestingly, the integrable structures underlying another realization of this approach, i.e.\ a particular $so(5)_2$ non-Abelian anyon chain \cite{FiFF14}, have been found to be closely related to those of an integrable clock model, the $\mathbf{Z}_5$ Fateev-Zamolodchikov model \cite{FaZa82,Albe94}. Such relationships have been observed between other integrable IRF models and vertex models (or, in the present context, anyon chains and quantum spin chains) \cite{Pasq88,Finch13}.

Below we will provide some evidence that the connection between chains of anyons constructed from the $so(n)_2$ fusion category and a class of $n$-state clock models can be extend beyond the integrable points in the space of coupling constants and to general odd $n$.  In the following two sections we introduce the Hamiltonians for both the $n$-state clock models and the $so(n)_2$ anyon chains.  Both can be defined for given $n$ and integers $1\le \ell\le n$ coprime to $2n$.  In the clock model the parameter $\ell$ enumerates the primitive $n$-th root of unity appearing in the diagonal Potts spin operator. On the other hand, in the anyon chain, $\ell$ labels the (gauge inequivalent) $F$-moves which have recently been constructed for the $so(n)_2$ fusion category \cite{ArFT16}.  We show that both families of Hamiltonian operators have a $\mathbf{Z}_n\otimes \mathbf{Z}_2$ symmetry related to a relabelling of the $\mathbf{Z}_n$ spin basis and an automorphism of the $so(n)_2$ fusion rules, respectively.  In addition, there exists a class of unitary transformation relating pairs of labels $(n,\ell)$.  Depending on $n$ and $\ell$, this unitary relation has different consequences: it may imply the existence of several inequivalent models of the same type (i.e.\ $n$-state clock or $so(n)_2$ anyon models) but with different realizations of the local interactions which have a second $\mathbf{Z}_2$ symmetry on the space of coupling constants.  This is found to be the case for $n=5$ leading to a nearest neighbour $so(5)_2$ anyon chain with different continuum limit than the one considered in Refs.~\cite{Finch.etal14,FiFF14}.
In other cases, e.g.\ for $n=7$, models with different $\ell$ are unitary equivalent which means that each one of them maps out the complete parameter space of coupling constants.  Within this space we also identify points where the chains become integrable.

In Section~\ref{sec:PhasePortraits} we study the zero temperature phase diagram of the models for $n=3,5,7$ using a variational matrix product ansatz for their translationally invariant states in the thermodynamic limit.  For the integrable models we also analyze the low energy behaviour: using the Bethe ansatz solution of these models we compute their ground state energies and classify the excitations.  
From the finite size spectrum, obtained also from the Bethe ansatz or by numerical diagonalization of the Hamiltonian, we compute the lowest scaling dimensions.  This allows to identify the low energy effective description in terms of rational CFT with extended symmetries related to that of the lattice model.  

Many of these rational CFTs have central charge $c=1$. Hence, they only can be rational due to extended chiral symmetry algebras $\mathcal{W}\mathfrak{g}$. These are Casimir algebras of affine Kac-Moody algebras $\hat{\mathfrak{g}}$ related to Lie algebras of type $B_\ell\simeq SO(2\ell+1)$ or $D_\ell\simeq SO(2\ell)$. Naively, these extended chiral symmetry algebras get larger with increasing $n$. However, for the particular value $c=1$ of the central charge, additional null fields appear such that many of the generators of these $\mathcal{W}$-algebras become algebraically dependent. For example, the $\mathcal{W}D_\ell$-algebras are generated by fields of conformal weights $2,4,6,\ldots,2(\ell-1),\ell$ but at the particular central charge $c=1$, only the fields with the weights $2,4,\ell$ remain algebraically independent. Since all rational CFTs with central charge $c=1$ are classified, it is known that, e.g.\ every rational $c=1$ $\mathbb{Z}_2$ orbifold theory has a $\mathcal{W}$-algebra of type $\mathcal{W}(2,4,k)$ with $k$ half-integer or integer. Therefore, Casimir-type $\mathcal{W}$ algebras, which exist for generic values of the central charge, must collapse to these smaller ones at $c=1$. The case $k=\ell\in\mathbb{Z}$ corresponds to $\mathcal{W}D_\ell$, and the case $k=\ell+\frac12\in\mathbb{Z}+\frac12$ corresponds to $\mathcal{W}\mathcal{B}_{0,\ell}$.  The latter is an alternative $\mathcal{W}$-algebra for the $B_\ell$ series with an fermionic generator, realized from the Lie-superalgebra $\mathcal{B}_{(0,\ell)}\simeq OSp(1|2\ell)$). Thus, all cases of the Lie algebras $\mathfrak{g}={SO}(n)$ are covered.

We note that the observed spectral equivalence between the clock models and anyon chains only holds up to degeneracies.  Furthermore, the Hilbert space of the anyons can be decomposed into sectors labelled by conserved topological charges which correspond to different boundary conditions in the clock models.  Thus, the full spectrum of conformal dimensions of the underlying CFT is, in general, present only in the finite size spectrum of the anyon chains.

Some technical background on the construction of anyon chains and the analysis of the Bethe equations as well on the rational CFTs relevant to the models considered in this paper are presented in the appendices.
\section{The $\mathbf{Z}_{n}$ clock models}

\subsection{The general model}
We construct a family of quantum spin chains with an odd number $n$ of states
per site.  For each pair of integers $(n,\ell)$, $1\leq \ell \leq n$ and
$\gcd(\ell,2n)=1$, the global Hamiltonian for a chain of length $L$ acting on
the Hilbert space $[\C^{n}]^{\tp L}$ is defined by
\begin{equation}
  \label{eqHamFZ}
  \begin{aligned}
    \mathcal{H}_{(n,\ell)}(\boldsymbol{c}) 
    & = \sum_{j=1}^{L} \left\{ c_{0}I + \sum_{k=1}^{n-1} c_{k}\left[X_{j}^{k}
        + Z_{j}^{k}Z_{j+1}^{-k}\right]\right\} \,,
  \end{aligned}
\end{equation}
where $c_{k}=c_{n-k}\in\R$.  The local operators $X_j$, $Z_j$ are operators
acting non-trivially on the spin at site $j$ as
\begin{align*}
  X & = \ket{n}\bra{1} + \sum_{k=1}^{n-1} \ket{k}\bra{k+1}\,, & 
  Z & = \sum_{k=1}^{n} \mathrm{e}^{\frac{4i\ell k\pi}{n}} \ket{k}\bra{k}\,.
\end{align*}
The $\frac{n+1}{2}$ coupling constants $c_{k}$ span the parameter space of the
clock model (\ref{eqHamFZ}).  However, as we are free to normalise and shift
the Hamiltonian, the parameter space of
$\mathcal{H}_{(n,\ell)}(\boldsymbol{c})$ equals the surface of a
$\frac{n-1}{2}$-sphere.

\subsection{Symmetries and maps}
To discuss the symmetries and equivalences between the general $n$-state clock
models we first consider bijections from the set of integers $\{1,\dots,n\}$
to itself.  Specifically we define the unique maps $\nu_{\downarrow}$ and
$\nu_{-}$
\begin{align*}
  \nu_{\downarrow}(k) & = k-1 \mod \, n\,,
  & \nu_{-}(k) & = - k \mod \, n\,,
\end{align*}
where $k\in\{1,\dots,n\}$. From these we can construct transformations of the
global Hamiltonian (\ref{eqHamFZ})
\begin{align*}
  U^{\nu} & = u^{\nu} \tp u^{\nu} \tp \cdots \tp u^{\nu}\,, \qquad
  u^{\nu} = \sum_{k=1}^{n} |\nu(k)\rangle\langle k|\,, 
\end{align*}
where $\nu$ is one of the aforementioned maps. 
It is straightforward to see that the Hamiltonian
$\mathcal{H}_{(n,\ell)}(\boldsymbol{c})$ is invariant under
$U^{\nu_{\downarrow}}$ and $U^{\nu_{-}}$ (we identify $c_{0}\equiv c_{n}$):
\begin{align*}
  U^{\nu_{\downarrow}} \mathcal{H}_{(n,\ell)}(\boldsymbol{c}) & =
  \mathcal{H}_{(n,\ell)}(\boldsymbol{c}) U^{\nu_{\downarrow}}\,,\\ 
  U^{\nu_{-}} \mathcal{H}_{(n,\ell)}(\boldsymbol{c}) & =
  \mathcal{H}_{(n,\ell)}(\boldsymbol{c}) U^{\nu_{-}}\,. 
\end{align*}
The first of these equations establishes the $\mathbf{Z}_{n}$ symmetry of the clock
model as a consequence of $u^{\nu_{\downarrow}}=X$.  The invariance under
$U^{\nu_{-}}$ together with the Hermiticity of the Hamiltonian, i.e.\
$c_{n-k}=c_{k}$, implies that the clock model is $\mathbf{Z}_{2}$-invariant.

\begin{figure}[ht]
\begin{center}
\begin{tikzpicture}
	\tikzstyle{every loop}=[]
	%
	\node (F21) at ( 0, 0) {$\mathcal{H}_{(5,1)}$};
	\node (F23) at ( 2, 0) {$\mathcal{H}_{(5,3)}$};loop
	\draw  [thick, ->, loop,looseness=5] (F21) to node [above] {\small $\nu_{3}$} (F21);
	\draw  [thick, ->, loop,looseness=5] (F23) to node [above] {\small $\nu_{3}$} (F23);
	\node (F21) at ( 6, -0.73) {$\mathcal{H}_{(7,1)}$};
	\node (F23) at ( 7.5, 1.85) {$\mathcal{H}_{(7,3)}$};
	\node (F25) at ( 9, -0.73) {$\mathcal{H}_{(7,5)}$};
	\draw  [thick, ->, transform canvas={yshift=0.1cm}] (F21) to node [above] {\small $\nu_{3}$} (F25);
	\draw  [thick, ->, transform canvas={xshift=-0.1cm}] (F25) to node [left, transform canvas={xshift=0.1cm,yshift=-0.2cm}] {\small $\nu_{3}$} (F23);
	\draw  [thick, ->, transform canvas={xshift=0.1cm}] (F23) to node [right, transform canvas={xshift=-0.1cm,yshift=-0.2cm}] {\small $\nu_{3}$} (F21);
	\draw  [thick, ->, transform canvas={xshift=-0.1cm}] (F21) to node [left] {\small $\nu_{5}$} (F23);
	\draw  [thick, ->, transform canvas={xshift=0.1cm}] (F23) to node [right] {\small $\nu_{5}$} (F25);
	\draw  [thick, ->, transform canvas={yshift=-0.1cm}] (F25) to node [below] {\small $\nu_{5}$} (F21);
\end{tikzpicture}
\end{center}
\caption{One can draw directed lines between the general Hamiltonians for each
  map $\nu_{t}$ (\ref{map-nut}).  For $n=5$ we see that the two Hamiltonians
  get mapped to themselves, indicating both have a $\mathbf{Z}_{2}$ symmetry in their
  phase space.  On the other hand, for $n=7$ the three Hamiltonians get mapped
  to each other, indicating the three different models are actually
  equivalent. \label{fig:hamrel}}
\end{figure}
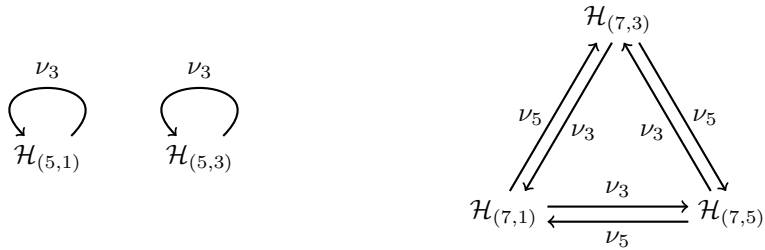

Similarly, we find that under the transformation generated from the bijections
$\nu_t$, $1\leq t \leq n$ and $\gcd(t,2n)=1$,
\begin{align}
  \label{map-nut}
  & \nu_{t}(k) =  tk \mod \, n\,,
\end{align}
the Hamiltonians $\mathcal{H}_{(n,\ell)}$ and $\mathcal{H}_{(n,\ell')}$ with
$\ell'=\pm t^{2}\ell\mod\,n$ are related as
\begin{align}
  \label{inv-nut}
  \mathcal{H}_{(n,\ell')}(\nu_{t}({\boldsymbol{c}})) & =
  \left[U^{\nu_{t}}\right]^{-1} \mathcal{H}_{(n,\ell)}({\boldsymbol{c}})\,
  U^{\nu_{t}}\,, \qquad 
  \mbox{where} \quad \nu_{t}(c)_{k} = c_{\nu_{t}(k)}\,.
\end{align}
This identity implies that some of the Hamiltonians (\ref{eqHamFZ}) for
different $(n,\ell)$ may be equivalent up to a basis transformation and
simultaneous permutation of the coupling constants.
Furthermore, we have $\ell'=\ell$ if $t^2=\pm1\mod n$.  This implies another
$\mathbf{Z}_{2}$-symmetry in the space of coupling constants as
$\nu_{t}\circ\nu_{t}=\mbox{id}$, see Figure~\ref{fig:hamrel}.

\subsection{Integrability}
Upon fine-tuning of the coupling constants the $\mathbf{Z}_{n}$ clock models
(\ref{eqHamFZ}) are integrable, i.e.\ members of a family of commuting
operators.  We find that there are two different types of such integrable
points:

For $c_k\equiv c$ independent of $k$, the clock model is the $n$-state Potts
model and the Hamiltonian can be given as \cite{Levy91}
\begin{equation}
  \mathcal{H}^{TL}_{(n)} \propto  \sum_{i=1}^{2L} E_{i}\,,\qquad
  E_{i} = \frac{1}{\sqrt{n}}\sum_{k=1}^{n-1} \,
  \begin{cases}
    X_{j}^{k} & i=2j-1 \\
    Z_{j}^{k}Z_{j+1}^{-k} & i=2j
  \end{cases}
  \,,
\end{equation}
where the $E_i$, $1 \leq i \leq 2L$, satisfy the relations of the periodic
Temperley--Lieb algebra
\begin{equation}
  \label{TLalgebra}
  \begin{aligned} 
    E_{i}E_{i} & = n E_{i}\,,\\
    E_{i}E_{i\pm1}E_{i} & = E_{i}\,, \\
    E_{i}E_{j} & = E_{j}E_{i}\,,
  \end{aligned}
\end{equation}
with $i -j \neq \pm 1$ (the indices and their difference are to be
interpreteded modulo $n$).  Clearly, these Hamiltonians are independent of the
parameter $\ell$.

The second type of integrable points of (\ref{eqHamFZ}) are the $\mathbf{Z}_{n}$
Fateev--Zamolodchikov models \cite{FaZa82}: 
for each pair of integers $(n,\ell_{FZ})$ with $1\le \ell_{FZ} \le n$ and $\gcd(\ell_{FZ},2n)=1$ a Fateev--Zamolodchikov $R$-matrix can be defined as
\begin{equation}
  \label{eqRmatFZ}
  \begin{aligned}
    R(u) 
    & = \sum_{a,b,c,d=1}^{n} \overline{W}_{b-c}(u) W_{b-d}(u)
    \overline{W}_{a-d}(u) W_{a-c}(u)
    \Big(\ket{a}\bra{b}\Big) \tp
    \Big(\ket{c}\bra{d}\Big)\,, 
  \end{aligned}
\end{equation}
where the weights are given by
\begin{align*}
  W_{a}(u) & = \prod_{k=1}^{g_{1}(a)}\sinh\left(\frac{i\pi (2k-1)
      \ell_{FZ}}{2n}+u\right) \prod_{k=g_{1}(a)+1}^{\frac{n-1}{2}}
  \sinh\left(\frac{i\pi (2k-1) \ell_{FZ}}{2n}-u\right)\,, \\ 
  \overline{W}_{a}(u) & = \prod_{k=1}^{g_{1}(a)} \sinh\left(\frac{i\pi (k-1)
      \ell_{FZ}}{n}-u\right) \prod_{k=g_{1}(a)+1}^{\frac{n-1}{2}}
  \sinh\left(\frac{i\pi k \ell_{FZ}}{n}+u\right)\,.
\end{align*}
Here and for later convenience we have defined the two functions
\begin{equation}
  \label{eqggmap}
  \begin{aligned}
    g_{1}:\Z & \rightarrow \{0,\dots,\tfrac{n-1}{2}\} & \mbox{such that} &&
    g_{1}(i) & = \pm i\mod\,\, n\,, \\ 
    g_{2}:\Z & \rightarrow \{0,\dots,n-1\} & \mbox{such that} &&
    g_{2}(i) & = \pm i\mod\,\, 2n\,. 
  \end{aligned}
\end{equation}

The $R$-matrices (\ref{eqRmatFZ}) satisfy the Yang--Baxter equation
\begin{align*}
  R_{12}(u)R_{13}(u+v)R_{23}(v) & = R_{23}(v)R_{13}(u+v)R_{12}(u)\,
\end{align*}
allowing for the construction of commuting transfer matrices 
\begin{align*}
  t(u) = \mbox{Tr}_{0} \left[ R_{01}(u)R_{02}(u)\cdots R_{0L}(u)\right]\,.
\end{align*}
The logarithmic derivative of the latter yields the integrable Hamiltonians
\begin{align*}
  \mathcal{H}^{FZ}_{(n)}(\ell_{FZ},J)
  &\propto J \left.\frac{\partial}{\partial u}\ln t(u)\right|_{u=0}\,,
\end{align*}
with $J=\pm1$.  Identifying 
\begin{equation}
  \label{eqellFZ}
  \ell_{FZ} = g_{2}(n - 2\ell t^2)\,
\end{equation}
these Hamiltonians coincide, apart from a constant shift, with the generic $n$-state clock model (\ref{eqHamFZ}) or its images (i.e.\ up to reordering of the basis) under the transformation $U^{\nu_t}$, Eq.\ (\ref{inv-nut}), i.e.\
\begin{align*}
  \mathcal{H}^{FZ}_{(n)}(\ell_{FZ},J)
  & = \sum_{j=1}^{L} \left\{ \sum_{k=1}^{n-1} \frac{J}{\sin\left(\frac{\pi
          k\ell_{FZ}}{n}\right)} \left[X_{j}^{kt} +
      Z_{j}^{kt}Z_{j+1}^{-kt}\right]\right\} 
  \\
  & \quad + L \left\{J\,\sum_{k=1}^{\frac{n-1}{2}} \frac{\sin\left(\frac{\pi
          \ell_{FZ}}{2n}\right)}{\cos\left(\frac{\pi k
          \ell_{FZ}}{n}\right)\cos\left(\frac{\pi (2k-1)
          \ell_{FZ}}{2n}\right)} - \frac{2}{\sin\left(\frac{\pi
          k\ell_{FZ}}{n}\right)} \right\} \mathbb{I}\,.
\end{align*}
(Note that the operators $Z_j$ depend explicitly on the root of unity parameter $\ell$.)

The $R$-matrix (\ref{eqRmatFZ}) is actually the uniform square limit of the
Fateev--Zamolodchikov $R$-matrix with a general root of unity (parametrized by
$\ell_{FZ}$), rather than the one presented originally \cite{FaZa82}.  It is
also a self-dual case of the chiral Potts $R$-matrix \cite{BaPA88}.  This
allows to make use of the functional relations from Ref.~\cite{BaSt90} to
express the transfer matrix eigenvalues in terms of the $d=(n-1)L$ roots
$\{u_j\}$ (some of which may be at $\pm\infty$) to the Bethe equations
\begin{equation}
  \label{baeFZ}
  \begin{aligned}
    \left(i\frac{\sinh\left(u_{j}+\frac{i\pi
            \ell_{FZ}}{4n})\right)}{\sinh\left(u_{j}-\frac{i\pi
            \ell_{FZ}}{4n}\right)}\right)^{2L} & = -\prod_{k=1}^{d}
    \left(\frac{\sinh\left(u_{j}-u_{k}+\frac{i\pi}{2}
          -\frac{i\pi\ell_{FZ}}{2n}\right)}{\sinh\left(u_{j}-u_{k}
          -\frac{i\pi}{2}+\frac{i\pi \ell_{FZ}}{2n}\right)}\right) \,.
  \end{aligned}
\end{equation}
In terms of the Bethe roots the energy and momentum of the corresponding state are given as
\begin{equation}
  \label{specFZ}
  \begin{aligned}
    E & =  iJ\left\{ \sum_{j=1}^{d} \frac{\cosh(u_{j}-\frac{i\pi
          \ell_{FZ}}{4n})}{\sinh(u_{j}-\frac{i\pi \ell_{FZ}}{4n})} \right\},
    \\
    P & = \mbox{Re}\left[ \frac{2}{i} \sum_{j=1}^{d}
      \ln\left[\sinh\left(-u_{j}+\frac{i\pi \ell_{FZ}}{4n}\right)\right]\right] +
    \mbox{const}. 
  \end{aligned}
\end{equation}

Clearly, the dynamics of the integrable Hamiltonians depends only upon the triple of parameters $(n,\ell_{FZ},J)$.  In Section~\ref{sec:PhasePortraits} and Appendix~\ref{app:thermo} below we shall classify the Bethe root configurations and discuss the low energy properties of the integrable clock models in greater detail, also see Ref.~\cite{Albe92} for $\ell_{FZ}=1$.  Here we note that, for $n=3$, the Temperley--Lieb and Fateev--Zamolodchikov integrable points coincide.

\section{The $so(n)_{2}$ anyon chains}
\subsection{The general model}

The $so(n)_{2}$ fusion category consists of objects
$\mathcal{I}=\{\epsilon_{\pm},\sigma_{\pm},\phi_{1},\dots,\phi_{p}\}$, $n=2p+1$. In contrast to the discussion in Appendix~\ref{app:fuscat} we do not distinguish between an object and its label.  The fusion rules for this category are
\begin{equation}
  \label{son-fusrules}
  \begin{aligned}
    \epsilon_{a} \tp \epsilon_{b} & \cong \epsilon_{ab}\,, 
    \qquad \epsilon_{a} \tp \sigma_{b} \cong \sigma_{ab} \,,
    \qquad \epsilon_{a} \tp \phi_{i} \cong \phi_{i}\,,
    \\
    \sigma_{a} \tp \sigma_{b} & \cong \epsilon_{ab} \oplus
    \bigoplus_{i=1}^{\frac{n-1}{2}} \phi_{i}\,,
    \qquad \sigma_{a} \tp \phi_{i} \cong \sigma_{+} \oplus \sigma_{-}\,,
    \\
    \phi_{i} \tp \phi_{j} & \cong \begin{cases}
      \epsilon_{+} \oplus \epsilon_{-} \oplus \phi_{g_{1}(2i)} & i=j \\
      \phi_{g_{1}(i-j)} \oplus \phi_{g_{1}(i+j)} & i\ne j
    \end{cases}\,,
  \end{aligned}
\end{equation}
where $a,b\in\{+,-\}$, $i,j\in\{1...\tfrac{n-1}{2}\}$.  We see that
$\epsilon_{+}$ is the identity object.

Compatible $F$- and $R$-moves have been constructed by Ardonne \emph{et al.} \cite{ArFT16}. It was found that there exists a family of gauge inequivalent $F$-moves labelled by the pairs $(\ell,\kappa)$ where $\kappa=\pm1$ and $1\leq \ell \leq n$ with $\gcd(\ell,2n)=1$. One can extract the parameters from the $F$-moves by considering the quantities
\begin{align*}
  \kappa & = \sqrt{n}\, \left(F^{\sigma_{+}\sigma_{+}\sigma_{+}}_{\sigma_{+}}\right)^{\epsilon_{+}}_{\epsilon_{+}} \left(F^{\sigma_{+}\epsilon_{+}\sigma_{+}}_{\epsilon_{+}}\right)^{\sigma_{+}}_{\sigma_{+}}, & 
  \ell & = \frac{n}{\pi} {\arccos}\left(\sqrt{\frac{\sqrt{n}}{2\kappa}\left(F^{\sigma_{+}\sigma_{+}\sigma_{+}}_{\sigma_{+}}\right)^{\phi_{1}}_{\phi_{1}} \left(F^{\sigma_{+}\phi_{1}\sigma_{+}}_{\phi_{1}}\right)^{\sigma_{+}}_{\sigma_{+}}}\right),
\end{align*}
where $\ell$ is the odd integer resulting from the expression.  For each of these sets of $F$-moves we can construct a set of projection operators.  We observe that every projection operator is independent of the choice of $\kappa$.  Therefore, without loss of generality we set $\kappa=+1$ for the remainder of the paper and denote the corresponding $F$-moves and projection operators $F(\ell)$ and $p(\ell)$, respectively.

The objects of the $so(n)_2$ fusion category can be grouped according to their quantum dimensions, i.e. the asymptotic contribution of a single such object to a large collection thereof: the sets $\{\epsilon_{\pm}\}$, $\{\phi_{i}\}_{i}$, and $\{\sigma_{\pm}\}$ contain particles of dimension $1$, $2$ and $\sqrt{n}$, respectively.
Following the general prescription above it is possible to construct uniform chains of each of these particles. Among these the $\epsilon_{\pm}$-anyon chains are trivial with their Hamiltonians necessarily being proportional to the identity.  The $\phi_{i}$-anyon chains, on the other hand, can be mapped to the XXZ model as the subcategory with objects $\{\epsilon_{\pm}, \phi_{i}\}_{i}$ is isomorphic to the category of representations of the group algebra $\C D_{n}$.  

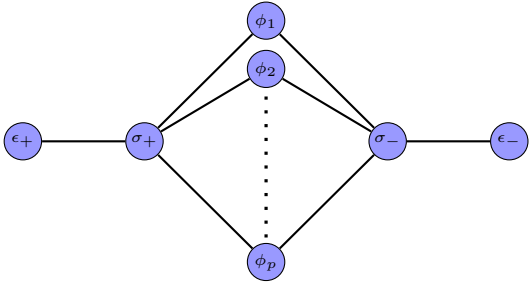
\begin{figure}[t]
\begin{center}{\tiny
\begin{tikzpicture}[scale=0.8]
	\tikzstyle{every node}=[circle,draw,thin,fill=blue!40,minimum size=14pt,inner sep=0pt]
	\tikzstyle{every loop}=[]
	\draw[loosely dotted,very thick] (0,1.2) -- (0,-2.0);
	\node (nep) at (-4, 0) {$\epsilon_{+}$};
	\node (nen) at ( 4, 0) {$\epsilon_{-}$};
	\node (nsp) at (-2, 0) {$\sigma_{+}$};
	\node (nsn) at ( 2, 0) {$\sigma_{-}$};
	\node (np1) at ( 0, 2) {$\phi_{1}$};
	\node (np2) at ( 0, 1.2) {$\phi_{2}$};
	\node (npp) at ( 0, -2) {$\phi_{p}$};
	\foreach \from/\to in
        {nep/nsp,nen/nsn,nsp/np1,np1/nsn,nsp/np2,np2/nsn,nsp/npp,npp/nsn}
        \draw[thick] (\from) -- (\to); 
\end{tikzpicture}}
\end{center}
\caption{Each state $\ket{\bf{a}}$ belonging to the set
  $\mathcal{B}_L^{(\sigma_{+})}$ can be interpreted as closed walk of length
  $2L$ on this graph. This is analogous to known models defined by walks on
  Dynkin diagrams \cite{Pasq87a}.\label{fig:adjgraph}}
\end{figure}

This leaves the $\sigma_{\pm}$-anyon chains.  It will become apparent below
that there is a mapping between the $\sigma_{+}$- and $\sigma_{-}$-anyon
chains.  Therefore we concern ourselves only with the former.
The Hilbert space of the $\sigma_{+}$-anyon chain is spanned by the vectors in
the set $\mathcal{B}_L^{(\sigma_{+})}$ (\ref{anybasis}).  An equivalent
definition of the set $\mathcal{B}_L^{(\sigma_{+})}$ is the all the closed
walks of length $2L$ on the (undirected) graph displayed in
Fig.~\ref{fig:adjgraph}.
The generic one-dimensional nearest-neighbour Hamiltonian for these $so(n)_2$
$\sigma_+$-anyons associated with the $F(\ell)$-moves is defined as
\begin{align}
  \label{eqHamSO} 
  \mathcal{H}_{(n,\ell)}({\boldsymbol\alpha}) & = \sum_{i=1}^{2L}
  \left[\alpha_{\epsilon_{+}} p_{i}^{(\epsilon_{+})}(\ell) +
    \sum_{k=1}^{\frac{n-1}{2}} \alpha_{\phi_{k}}\, p_{i}^{(\phi_{k})}(\ell)\right].
\end{align}
Without affecting the dynamics, one can fix $\alpha_{\epsilon_{+}}$ and have
the remaining coupling constants $\alpha_{\phi_{k}}$ as co-ordinates on the
surface of a $\tfrac{n-1}{2}$-sphere.

\subsection{Symmetries and equivalent $so(n)_{2}$ models}
As stated in Appendix~\ref{app:fuscat} global symmetries and maps between
equivalent anyon models can be inferred from the automorphisms of the fusion
rules and the associated monoidal equivalences between $F$-moves.  Along with
the $F$-moves, Ardonne \emph{et al.} \cite{ArFT16} also discussed the monoidal
and gauge equivalences of $so(n)_{2}$.  The automorphisms of $so(n)_{2}$
fusion rules are given by the sets
\begin{align*}
  \left\{ \nu_{\pm} \circ \nu_{t}\, |\, 
     1 \leq t \leq n, \, \gcd(t,2n)=1 \right\}\,.
\end{align*}
The non-trivial actions of $\nu_{\pm}$ and $\nu_{t}$ are
\begin{equation}
  \label{maps-son}
  \begin{aligned}
    \nu_{a}(\sigma_{b}) & = \sigma_{ab}\,, \quad a,b\in\{+,-\}\,,\\
    \nu_{t}(\phi_{i})   & = \phi_{g_{1}(ti)}\,, \quad
    i\in{1...\tfrac{n-1}{2}}\,,
  \end{aligned}
\end{equation}
where $g_{1}$ is the map defined in Eq.~(\ref{eqggmap}).

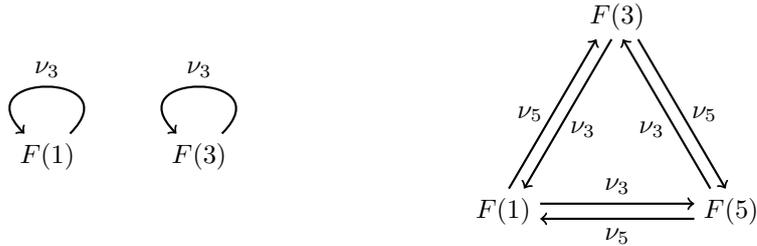
\begin{figure}[t]
\begin{center}
\begin{tikzpicture}
	\tikzstyle{every loop}=[]
	%
	\node (F21) at ( 0, 0) {$F({1})$};
	\node (F23) at ( 2, 0) {$F({3})$};loop
	\draw  [thick, ->, loop,looseness=5] (F21) to node [above] {\small $\nu_{3}$} (F21);
	\draw  [thick, ->, loop,looseness=5] (F23) to node [above] {\small $\nu_{3}$} (F23);
	\node (F21) at ( 6, -0.73) {$F({1})$};
	\node (F23) at ( 7.5, 1.85) {$F({3})$};
	\node (F25) at ( 9, -0.73) {$F({5})$};
	\draw  [thick, ->, transform canvas={yshift=0.1cm}] (F21) to node [above] {\small $\nu_{3}$} (F25);
	\draw  [thick, ->, transform canvas={xshift=-0.1cm}] (F25) to node [left, transform canvas={xshift=0.1cm,yshift=-0.2cm}] {\small $\nu_{3}$} (F23);
	\draw  [thick, ->, transform canvas={xshift=0.1cm}] (F23) to node [right, transform canvas={xshift=-0.1cm,yshift=-0.2cm}] {\small $\nu_{3}$} (F21);
	\draw  [thick, ->, transform canvas={xshift=-0.1cm}] (F21) to node [left] {\small $\nu_{5}$} (F23);
	\draw  [thick, ->, transform canvas={xshift=0.1cm}] (F23) to node [right] {\small $\nu_{5}$} (F25);
	\draw  [thick, ->, transform canvas={yshift=-0.1cm}] (F25) to node [below] {\small $\nu_{5}$} (F21);
\end{tikzpicture}
\end{center}
\caption{One can draw directed lines between the sets of $F$-moves to indicate
  one set of $F$-moves is related/mapped to another via an automorphism
  $\nu_{t\neq1}$. For $so(5)_{2}$ (left) we set that there are two sets of
  monoidally equivalent $F$-moves indicating
  $\mathcal{H}_{(5,1)}(\boldsymbol{\alpha})$ and
  $\mathcal{H}_{(5,3)}(\boldsymbol{\alpha})$ are inequivalent and their
  parameter spaces must possess a $\mathbf{Z}_{2}$ symmetry. For $so(7)_{2}$ (right)
  we see that the $F$-moves are monoidally equivalent and thus
  $\mathcal{H}_{(7,\ell)}(\boldsymbol{\alpha})$, $\ell=1,3,5$ are
  equivalent. This is consistent with the relationships shown in Fig. \ref{fig:hamrel}. \label{fig:monrel}}
\end{figure}

The $F$-moves labelled $(\ell,\kappa)$ are monoidally related to themselves
via the map $\nu_{-}$.  Thus there is a one-to-one equivalence between
$\sigma_{+}$-anyon and $\sigma_{-}$-anyon chains, as asserted above.
The automorphism $\nu_{t}$ relates the $F({\ell})$ to $F({\ell'})$ if and only
if $\ell'=g_{2}(t^{2}\ell)$ where $g_{2}$ is the map defined in
Eq.~(\ref{eqggmap}).  
Thus we are able to completely determine the mappings between Hamiltonians and
the internal symmetries of the Hamiltonians.  However, since the number of
monoidally inequivalent $F$-moves, and consequently inequivalent Hamiltonians,
depends on $n$ not much can be said about the mappings and symmetries in
general.  Nevertheless, three interesting observations can be made:
\begin{enumerate}
\item If $\nu_{t}$ gives rise to an internal symmetry of the parameter spaces
  of $\mathcal{H}_{(n,\ell)}$, it must also give rise to an internal symmetry
  of the parameter spaces of $\mathcal{H}_{(n,\ell')}$, independent of
  $\ell'$: this follows from the invertibility of $\ell$ and $\ell'$ mod $2n$.

\item The internal symmetries of the parameter spaces are $\mathbf{Z}_{2}$ symmetries:
  if $\nu_{t\neq1}$ gives rise to an internal symmetry of the parameter spaces
  of $\mathcal{H}_{(n,\ell)}$ then it follows that $g_{2}(t^{2})=1$,
  $g_{1}(t^{2})=1$ and lastly $\nu_{t}\circ\nu_{t}=\mbox{id}$.

\item The total area of inequivalent points for the $so(n)_{2}$ Hamiltonians
  $\mathcal{H}_{(n,\ell)}$ equals the surface of a $\tfrac{n-1}{2}$-sphere:
  there are the same number of Hamiltonians $\mathcal{H}_{(n,\ell)}$ as
  automorphisms $\nu_{t}$ and every $\nu_{t}$ is unique.
\end{enumerate} 
These observations allow us to make inferences. For instance, for $so(5)_{2}$,
the two Hamiltonians, $\mathcal{H}_{(5,1)}$ and $\mathcal{H}_{(5,3)}$, will
either be equivalent or the parameter spaces will have a $\mathbf{Z}_{2}$ symmetry,
with the latter is found to be true (see Figure \ref{fig:monrel}).
In contrast, for $so(7)_{2}$, the three Hamiltonians $\mathcal{H}_{(7,\ell)}$,
$\ell=1,3,5$, must all be equivalent as there is no other possibility of
having the total area of inequivalent points equalling the surface of a
sphere.

\subsection{Integrability}
\subsubsection*{The general face model formulation}
In the vertex language solutions to the Yang-Baxter equation are often
associated with quasi-triangular Hopf algebras, typically quantum groups.  In
a similar fashion, in the face (or anyon) language one often identifies one or
more solution to the Yang--Baxter equation associated with a braided fusion
category. Using the basis defined earlier we define an $R$-matrix to be of the
form
\begin{align*}
  \bra{\bf{a}'} R_{i}(u) \ket{\bf{a}} & =
  \Wop[a_{i-1}][a_{i}'][a_{i}][a_{i+1}][u] \prod_{k\neq i}
  \delta_{a_{k}}^{a_{k}'}, \quad 
  \hbox{where}\quad   \ket{\bf{a}}=\ket{a_1,a_2,\cdots,a_{2L}}
  \in\mathcal{B}_L^{(j)}\,, 
\end{align*}
and must satisfy the face Yang--Baxter equation
\begin{align*}
  R_{i}(u)R_{i+1}(u+v)R_{i}(v) & = R_{i+1}(v)R_{i}(u+v)R_{i+1}(u).
\end{align*}
The SOS transfer matrix $t(u)$ is then defined by
\begin{align*}
  \bra{\bf{a}'} t(u) \ket{\bf{a}} & = \prod_{i=1}^{2L}
  \Wop[a_{i-1}'][a_{i}'][a_{i-1}][a_{i}][u-u_{i}]\,\quad
  \ket{\bf{a}},\ket{\bf{a}'}\in \mathcal{B}_L^{(j)}.
\end{align*}
This can be used to construct an integrable quantum Hamiltonian
\begin{align*}
  \mathcal{H} & \propto  \left. \frac{d}{du} \ln(t(u)) \right|_{u=0}.
\end{align*}
In the case that the $R$-matrix satisfies the regularity condition, i.e. $R(0)\propto I$, the Hamiltonian will describe a periodic chain of interacting particles with nearest-neighbour interactions.

\subsubsection*{The $so(n)_{2}$ integrable points}
In this section we identify pairs of integrable points associated with
solutions to the Yang--Baxter equation. The solutions of the Yang--Baxter
equation come in two types, either Temperley--Lieb $R$-matrices or $\mathbf{Z}_{n}$
Fateev--Zamolodchikov like $R$-matrices.

The Temperley--Lieb $R$-matrices are defined as
\begin{align*}
  R_{i}(u) 
  & = \frac{\sinh(\gamma-u)}{\sinh(\gamma+u)} p^{(\epsilon_{+})} + \left[ \sum_{k=1}^{\frac{n-1}{2}} p_{i}^{(\phi_{k})} \right]
\end{align*}
where $2\cosh(\gamma)=\sqrt{n}$.  A simple rescaling of the projection
operators onto the identity object, $E_i = \sqrt{n}\,p_i^{(\epsilon_{+})}$,
leads to a representation of the Temperley--Lieb algebra (\ref{TLalgebra}).
As a consequence the spectrum of the corresponding integrable points coincides
with that of the XXZ spin chain (up to boundary conditions).  Specifically,
this implies that the anyon model at the Temperley--Lieb integrable points
${\boldsymbol\alpha}=\pm(1,0,\dots,0)$ will be gapped for $n>4$.  These points
are fixed points of any symmetry generated by an automophism of the fusion
rules.  Moreover, at these two points the Hamiltonian
$H_{(n,\ell)}({\boldsymbol\alpha})$ is the same for all allowed $\ell$.

The second class of $R$-matrices we consider is connected to the $\mathbf{Z}_{n}$
Fateev--Zammolodchikov like $R$-matrix (\ref{eqRmatFZ}).  Analogously to the
integrable points for the $\mathbf{Z}_{n}$ clock model, for the $F$-moves $F({\ell})$
and automophism $\nu_{t}$ we define the parameter
\begin{align*}
  \ell_{FZ} & = g_{2}(n - 2\ell t^{2})\,.
\end{align*}
The $R$-matrices are defined to be
\begin{equation}
\begin{aligned}
  R_{i}^{(\ell)}(u) = &
  \left[\prod_{k=1}^{\frac{n-1}{2}}\sinh\left(u-\frac{i\pi
        (1-2k)\ell_{FZ}}{2n}\right)\right]\\
  & \times \left\{p_{i}^{(\epsilon_{+})}(\ell) + \sum_{j=1}^{\frac{n-1}{2}}
    \left[ \prod_{k=1}^{j} \frac{\sinh(u+\frac{i\pi
          (1-2k)\ell_{FZ}}{2n})}{\sinh(u-\frac{i\pi (1-2k)\ell_{FZ}}{2n})}
    \right] \, p_{i}^{\left(\phi_{g_{1}(2jt)}\right)}(\ell)\right\}
\end{aligned}
\end{equation}
These operators have been verified to satisfy the Yang--Baxter equation for $3\leq n \leq 9$ and conjectured for larger $n$. Their existence implies that the Hamiltonians (\ref{eqHamSO}) with the particular choice of coupling constants $\mathbf{\alpha}$ 
\begin{align*}
	\mathcal{H}_{(n,\ell)}
	& = -J\left\{ \sum_{i=1}^{2L} \sum_{j=1}^{\frac{n-1}{2}} 2\left[
            \sum_{k=1}^{j}
            \frac{\cos\left(\frac{\pi(2k-1)\ell_{FZ}}{2n}\right)}{
                  \sin\left(\frac{\pi(2k-1)\ell_{FZ}}{2n}\right)}  
          \right] p^{\left(\phi_{g_{1}(2jt)}\right)}(\ell)\right\}  
	+ 2JL\left\{\sum_{j=1}^{\frac{n-1}{2}}
          \frac{\cos\left(\frac{\pi(2j-1)\ell_{FZ}}{2n}\right)}{
                \sin\left(\frac{\pi(2j-1)\ell_{FZ}}{2n}\right)} 
        \right\} , 
\end{align*}
are integrable where $J=\pm1$. The Hamiltonian can be normalised to the form
given above, but for our purposes that is unnecessary.

Using the same approach as in \cite{FiFF14}, additional $R$-matrices labeled
by $b\in\mathcal{I}$ can be constructed starting from the given set of
$F$-moves.  The resulting transfer matrices are mutually commuting and satisfy
a set of fusion relations.  Within this construction, the topological charges
$Y_b$, defined in Eq.~(\ref{Ytopo}) for the general anyon model, are obtained
from these transfer matrices in the braiding limit $u\to\infty$.

Analyticity arguments imply that the transfer matrix eigenvalues can be
parametrized by a set of complex numbers $\{u_j\}$ solving the Bethe equations
\begin{align}
  \label{baeSO}
  \left(i\frac{\sinh\left(u_{j}+\frac{i\pi
          \ell_{FZ}}{4n})\right)}{\sinh\left(u_{j}-\frac{i\pi
          \ell_{FZ}}{4n}\right)}\right)^{2L} & = -s \prod_{k=1}^{d}
  \left(\frac{\sinh\left(u_{j}-u_{k}+\frac{i\pi}{2}-\frac{i\pi
          \ell_{FZ}}{2n}\right)}{\sinh\left(u_{j}-u_{k}-\frac{i\pi}{2}+\frac{i\pi
          \ell_{FZ}}{2n}\right)}\right) 
\end{align}
where $s=\pm1$ is the eigenvalue of the topological charge $Y_{\epsilon_-}$
\cite{FiFF14}.
We see that the Bethe equations coincide with those for Fateev--Zamolodchikov
$\mathbf{Z}_{n}$ clock model integrable points, Eq.~(\ref{baeFZ}), up to a twist in
the boundary conditions in the sector with $s=-1$.  This implies some
connection and we expect that the two different models share dynamics at these
integrable points once the topological sectors of the anyon chain are
identified with suitable boundary conditions for the clock model.

The energy and momentum are give by
\begin{equation}
  \label{specSO}
  \begin{aligned}
  E & =  iJ\left\{ \sum_{j=1}^{d} \frac{\cosh(u_{j}-\frac{i\pi \ell_{FZ}}{4n})}{\sinh(u_{j}-\frac{i\pi \ell_{FZ}}{4n})} \right\},  \\
  P & = \mbox{Re}\left[ \frac{1}{i} \sum_{j=1}^{d} \ln\left[\sinh(-u_{j}+\frac{i\pi \ell_{FZ}}{4n})\right]\right] + \mbox{const}.
\end{aligned}
\end{equation}
Note that the momentum for the anyon chain differs to the that coming from the
$\mathbf{Z}_{n}$ models due to the factor of two difference between the chain lengths
of the two models.

\subsection{Mapping between anyon and clock models}
The analysis of the symmetries and structures underlying their integrable points above has revealed striking similarities between the $so(n)_2$ anyon chains and the $\mathbf{Z}_n$ clock models -- in spite of the very different Hilbert spaces on which they are defined.
Here we put forward a more general relationship between the $\mathbf{Z}_{n}$ clock models for given parameter $\ell$ and chain length $L$ and the $so(n)_{2}$ anyons chain, built from the $F({\ell})$-moves with chain length $2L$.  Specifically, we claim that when
\begin{equation}
  \label{so2fz}
  \begin{aligned}
    c_{0} & = \frac{2}{n}\alpha_{\epsilon_{+}}
            + \frac{4}{n} \sum_{k=1}^{\frac{n-1}{2}} \alpha_{\phi_{k}}\,, \\
    c_{j} & = \frac{1}{n}\alpha_{\epsilon_{+}}
            + \frac{2}{n} \sum_{k=1}^{\frac{n-1}{2}}
              \cos\left(\frac{2\ell jk\pi}{n}\right)\alpha_{\phi_{k}}\,, 
    \qquad\mbox{for}\quad 1\leq j \leq \frac{n-1}{2}\,,
  \end{aligned}
\end{equation}
the two Hamiltonians $\mathcal{H}_{(n,\ell)}(\boldsymbol{c})$ and $\mathcal{H}_{(n,\ell)}(\boldsymbol{\alpha})$, defined by Equations (\ref{eqHamFZ}) and (\ref{eqHamSO}) respectively, share some energy levels (although their degeneracies may differ).  We have checked this conjecture numerically for small system sizes and found that, depending on the chain lengths (e.g.\ for $L$ being multiples of $4$) this includes the ground state.  
This relationship between models, along with the underlying algebraic structure of the anyon model, suggests the existence of a face-vertex (or anyon-spin) correspondence \cite{Pasq88,Finch13}.

\section{Phase portraits and low energy effective theories}
\label{sec:PhasePortraits}

In the following we study the phase diagrams of the $\mathbf{Z}_n$ clock and $so(n)_2$ anyon models for $n=3,5,7$ based on variational matrix product states (MPS) representing translationally invariant states of the chains in the thermodynamic limit, diagonalization of the Hamiltonian for small lattice sizes $L$ using the Lanczos algorithm, and -- for the integrable points -- numerical solution of the Bethe equations (\ref{baeFZ}) and (\ref{baeSO}) for chains of a few hundred sites.  For some of these models we are able to identify the conformal field theories describing the continuum limit based on the finite size scaling of the ground state and low lying excitations, i.e.\ \cite{BlCN86,Affl86}
\begin{equation}
  \label{fscft}
  \begin{aligned}
    E_0(L) &= L\epsilon_\infty - \frac{\pi}{6L} v^{(F)} c\,,\\
    E_n(L) &= E_0(L) + \frac{2\pi}{L} v^{(F)} X_n\,,&
    P_n(L) &= P_0(L) + \frac{2\pi}{L} s_n + \mathrm{const.}
\end{aligned}
\end{equation}
where $c$ is the central charge of the underlying Virasoro algebra and
$v^{(F)}$ is the velocity of massless excitations.  From the spectrum
of scaling dimensions $X_n=\left( h+\bar{h} \right)$ and conformal
spins $s_n=\left( h-\bar{h} \right)$ the operator content, i.e.\
primary fields with conformal weights $(h,\bar{h})$, can be
determined.

\subsection{The $\mathbf{Z}_{3}$ and $so(3)_{2}$ models}
The spectrum of low energy excitations of the $\mathbf{Z}_{3}$ clock model or, equivalently, the self-dual three-state critical Potts model, has been studied in Ref.~\cite{AlDM92,AlDM93}.  Its partition function has been computed both for the antiferromagnetic and the ferromagnetic case and found to reproduce the modular invariant partition function of a $\mathbf{Z}_4$ parafermion model with $c=1$, and that of the three-state Potts model (the minimal model $\mathcal{M}_{(5,6)}$) with central charge $c=\frac{4}{5}$, respectively \cite{KeMc93,DKMM94}.

\begin{figure}[t]
    \centering
    (a) \includegraphics[width=.45\linewidth]{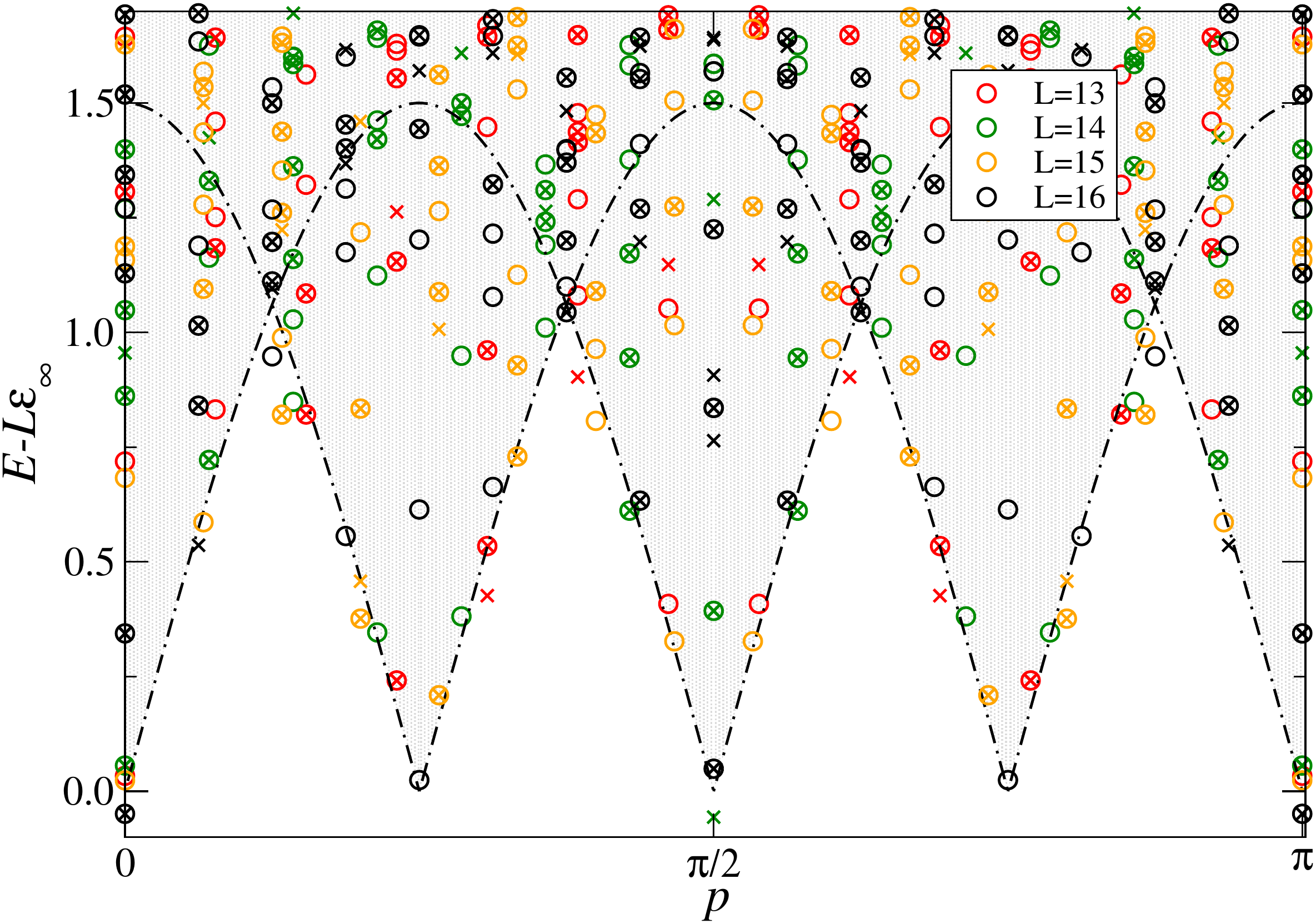}
    \hspace{\fill}
    (b) \includegraphics[width=.45\linewidth]{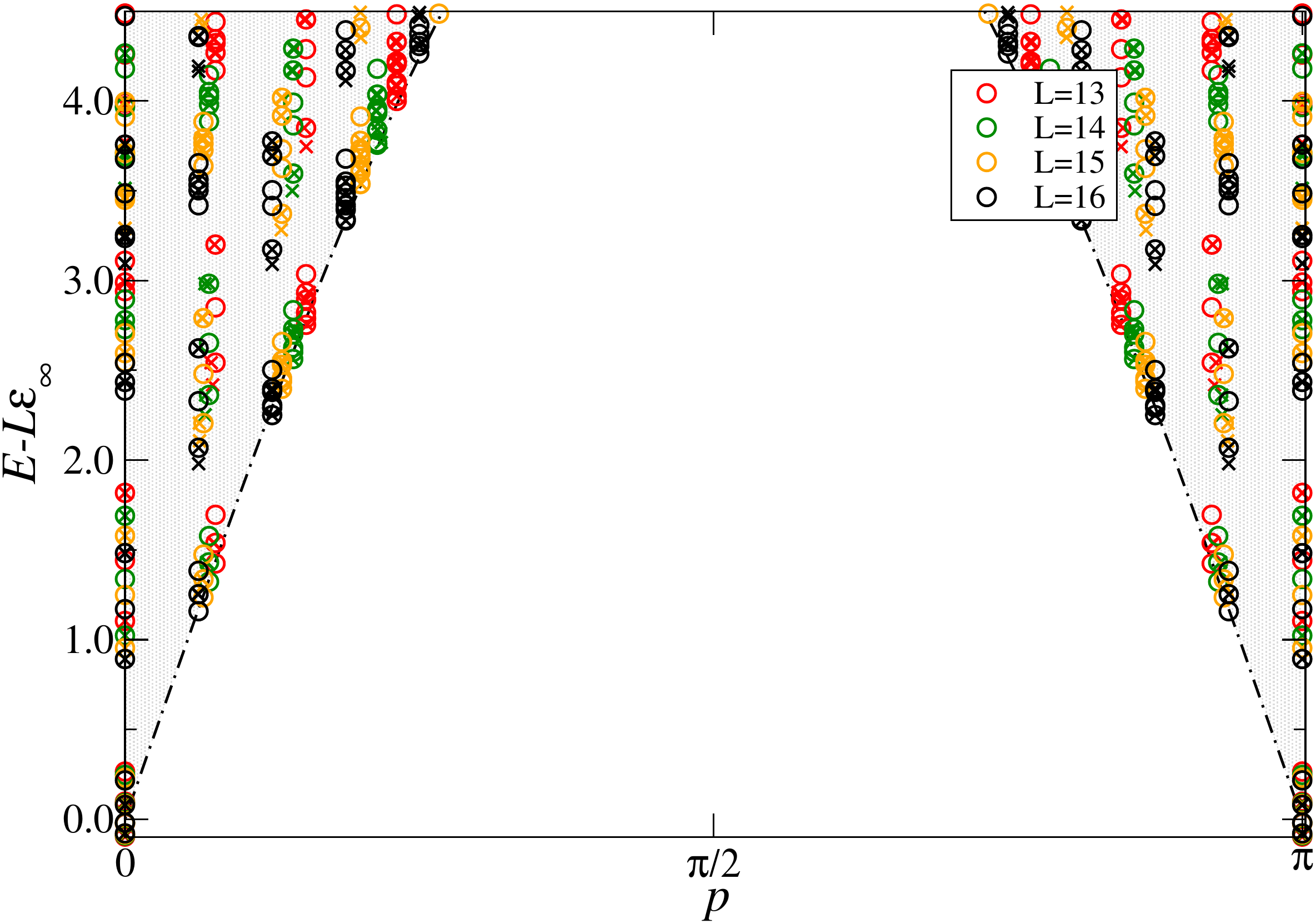}
    \caption{(a) Finite size spectra of the integrable $(3,1,+1)$ $\mathbf{Z}_3$
      clock model ($\times$) for system sizes $L=13$, $14$, $15$ and $16$
      (resp.\ the $so(3)_2$ anyon model ($\bigcirc$) with $\widetilde{L}=2L$
      sites).  (b) Same for the $(3,1,-1)$ models.
      Energies are measured relative to value obtained from the
      ground state energy density in the thermodynamic limit, momenta are
      scaled to the values taken in the anyon chain.  The shaded area
      indicates the continuum of excitations in the thermodynamic limit.}
    \label{fig:n3_spec}
\end{figure}

The $so(3)_{2}$ chain on the other hand is equivalent to a chain of $su(2)_4$ anyons.  This is a deformation of the spin-$1/2$ Heisenberg model \cite{FTLT07}, more commonly known as the $A_{5}$ restricted solid-on-solid model \cite{AnBF84,Pasq87a} which is another formulation of the three-state Potts model.
To illustrate the equivalence between the $\mathbf{Z}_{3}$ clock and $so(3)_{2}$ anyon models with coupling constants related by Eq.~(\ref{so2fz}) we plot the spectra of their low energy excitations in Fig.~\ref{fig:n3_spec}.

For the antiferromagnetic model, i.e.\ $(n,\ell_{FZ},J)= (3,1,+1)$, we have identified the Bethe ansatz solutions corresponding to the lowest of the conformal weights of $\mathbf{Z}_4$ parafermions (\ref{cft_specZ4}),
see Table~\ref{table:fsdata31p}.
This parafermion CFT is realized by the $\mathbf{Z}_2$ orbifold of a $U(1)$ boson compactified on a circle of radius $R^2=3/2$ \cite{Ginsparg88}.  We also note that this spectrum coincides with that of the $\mathcal{W}D_3(5,6)$ rational CFT.
\begin{table}[t] 
\begin{ruledtabular}
\begin{tabular}{ccccccc} 
  & extrapolation & $s$ & conjecture & $\Delta_{(1,+)}$ & $\Delta_{(1,-)}$ & comment \tabularnewline \hline
  c &  1.000000 &   & 1  & 0 & 0 & a,c \tabularnewline 
  X &  0.125000 & 0 & $(\frac{1}{16},\frac{1}{16})$ & 0 & 0 & a \tabularnewline 
    & 0.166667 & 0 & $(\frac{1}{12},\frac{1}{12})$ & $-1$ & $-1$ & a,c \tabularnewline 
    & 0.666667 & 0 & $(\frac{1}{3},\frac{1}{3})$ & $-2$ & 0 & a,c \tabularnewline 
\end{tabular}
\end{ruledtabular}
\caption{\label{table:fsdata31p}
  Central charge $c$ and conformal spectrum of the $(3,1,+1)$ model:
  shown are the extrapolated values for $c$ and the conformal dimensions 
  $X=h+\bar{h}$ together with the conformal spins $s=h-\bar{h}$ of the
  primaries as obtained from the Bethe ansatz solution.  The conjectured 
  values are the central charge  and pairs of conformal weights $(h,\bar{h})$ 
  for $\mathbf{Z}_4$ parafermions, Eq.~(\ref{cft_specZ4}).
  $\Delta_\gamma$ denote the difference in the number of $\gamma$-patterns
  in the Bethe root configuration as compared to the ground state, see
  {Appendix~\ref{app:thermo}}.  In the last column we indicate in which model (a: anyon chain, c: clock model) the corresponding level can be observed -- possibly subject to further constraints on the system size $L$.
} 
\end{table}

Similarly, for the ferromagnetic model, $(n,\ell_{FZ},J)=(3,1,-1)$, we have identified the Bethe ansatz solutions yielding the lowest of the conformal weights (\ref{cft_specPotts3}) of the minimal model $\mathcal{M}_{(5,6)}$, see Table~\ref{table:fsdata31m}.
\begin{table}[t] 
\begin{ruledtabular}
\begin{tabular}{cccccccc} 
  & extrapolation & $s$ & conjecture & $\Delta_{(1,+)}$ & $\Delta_{(1,-)}$ & $\Delta_{(2,+)}$ & comment \tabularnewline \hline
  c & 0.800000(1) &   & $\frac45$  & 0 & 0 & 0 & a,c \tabularnewline 
  X & 0.050003(1) & 0 & $(\frac{1}{40},\frac{1}{40})$ & 0 & 0 &  0 & a \tabularnewline 
    & 0.133337(1) & 0 & $(\frac{1}{15},\frac{1}{15})$ & 0 & 0 & $-1$ & a,c \tabularnewline 
    & 0.250000(1) & 0 & $(\frac{1}{8},\frac{1}{8})$ & $2$ & 0 & $-1$ & a \tabularnewline 
    & 0.801(1) & 0 & $(\frac{2}{5},\frac{2}{5})$ & - & - & -  & a,c\tabularnewline 
\end{tabular}
\end{ruledtabular}
\caption{\label{table:fsdata31m}
  Similar as Table~\ref{table:fsdata31p} but for the $(3,1,-1)$ model.  The conjectured values for the conformal weights are those from the critical three-state Potts model (\ref{cft_specPotts3}).
  The changes $\Delta_\gamma$ of root patterns have been omitted in the last excitation as deformations due to finite size obscure the configuration of the thermodynamic limit.
} 
\end{table}
We note that the conformal weights of the twist operators, $h\in\{\frac{1}{40}, \frac18, \frac{21}{40}, \frac{13}{8}\}$, corresponding to disorder operators in the Potts model are observed only in the low energy spectrum of anyon model.  They are absent in the clock model with periodic boundary conditions as considered here but do appear in the spectrum of the $\mathbf{Z}_3$ clock model for twisted boundary conditions \cite{Card86b}.

\subsection{The $\mathbf{Z}_{5}$ and $so(5)_{2}$ models}
The $so(5)_{2}$ model is the first model in which the parameter space is not a
single point. As $F({1})$ and $F({3})$ are not monodially equivalent there will
be two distinct models each with a global $\mathbf{Z}_{2}$ symmetry in the
Hamiltonian. Instead of using the couplings $\boldsymbol{\alpha}$ we switch to
polar coordinates with fixed radius. This gives the general Hamiltonians
\begin{equation}
  \label{hamil5}
  \mathcal{H}_{(5,\ell)}(\theta) = \sum_{j=1}^{L} \left[
    \cos\left(\theta+\frac{\pi}{4}\right) \, p_{j}^{(\phi_{1})}(\ell)
    + \sin\left(\theta+\frac{\pi}{4}\right) \, p_{j}^{(\phi_{2})}(\ell)\right]
\end{equation}
where $\ell=1,3$ and $\theta \in [0, 2\pi)$. The $\mathbf{Z}_{2}$ symmetry
manifests itself as
\begin{equation}
  \mathcal{H}_{(5,\ell)}(-\theta) = U^{-1}\mathcal{H}_{(5,\ell)}(\theta)U\,.
\end{equation}
At the TL points $\theta_{TL}=0,\pi$, i.e.\ the fixed points of this symmetry,
the Hamiltonians coincide.  In previous work on the model with $\ell=3$
\cite{Finch.etal14,FiFF14} the TL equivalence to the XXZ spin chain has been
employed to conclude that the excitations at $\theta=0$ are massive with a
tiny energy gap $\Delta E\simeq 2.910^{-4}$.  This gap is difficult to resolve
numerically but based on continuity arguments it is expected that this gapped
phase extends to small finite values of $\theta$.  Similarly, the spectrum at
$\theta=\pi$ is highly degenerate indicating a first order transition.

\begin{figure}[t]
\includegraphics[width=0.45\textwidth]{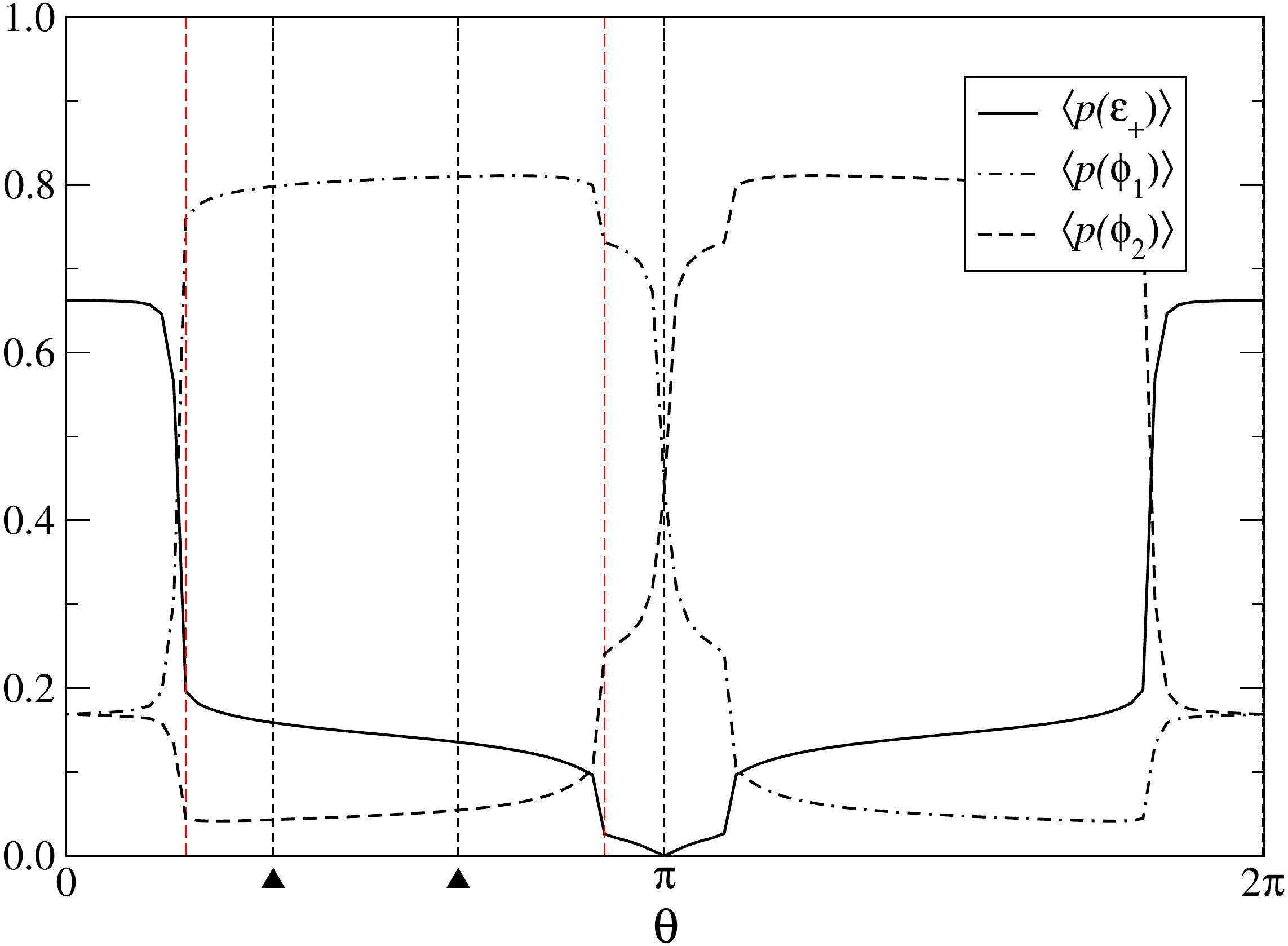}
\hspace*{\fill}
\includegraphics[width=0.45\textwidth]{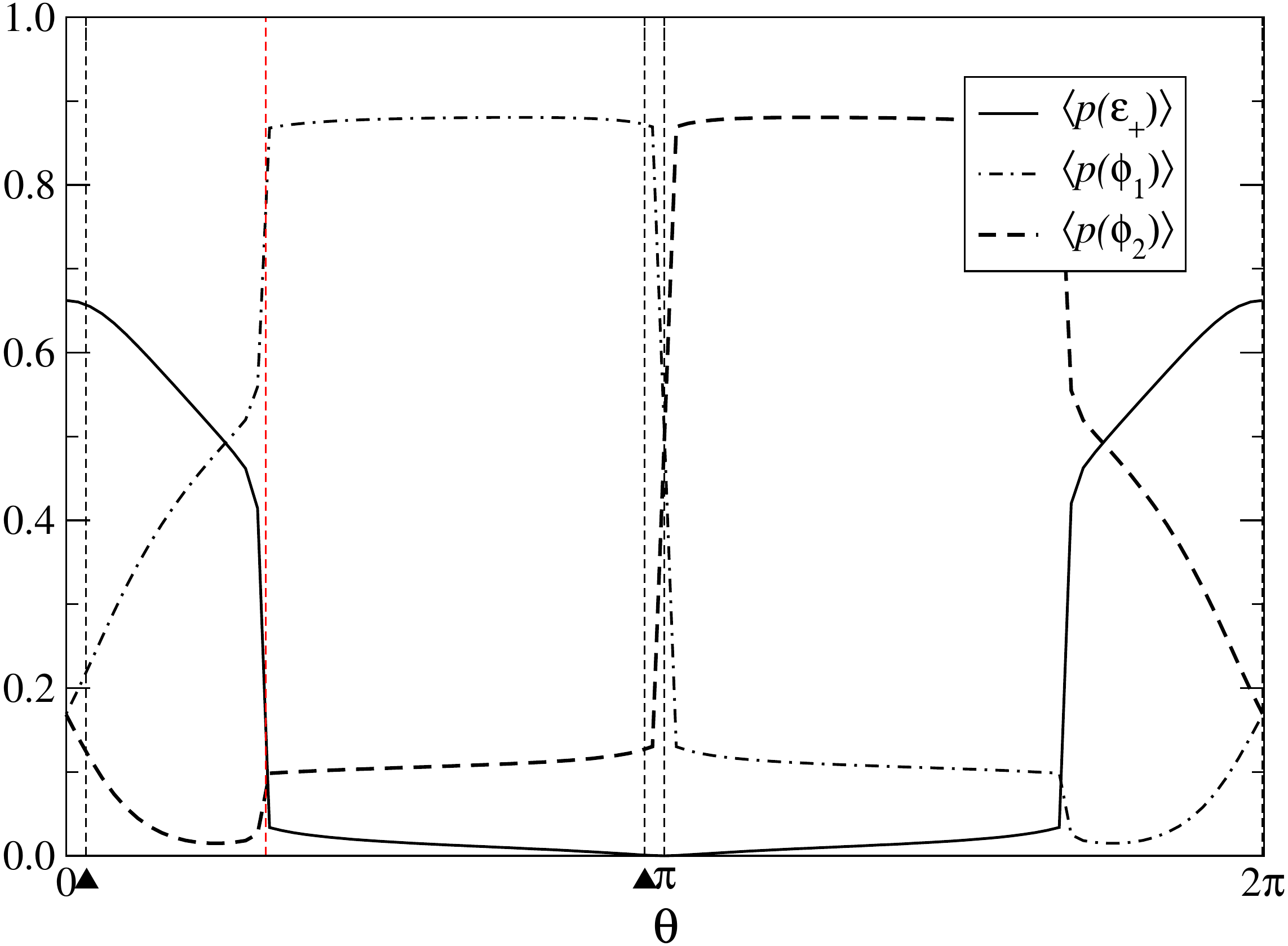}

\includegraphics[width=0.45\textwidth]{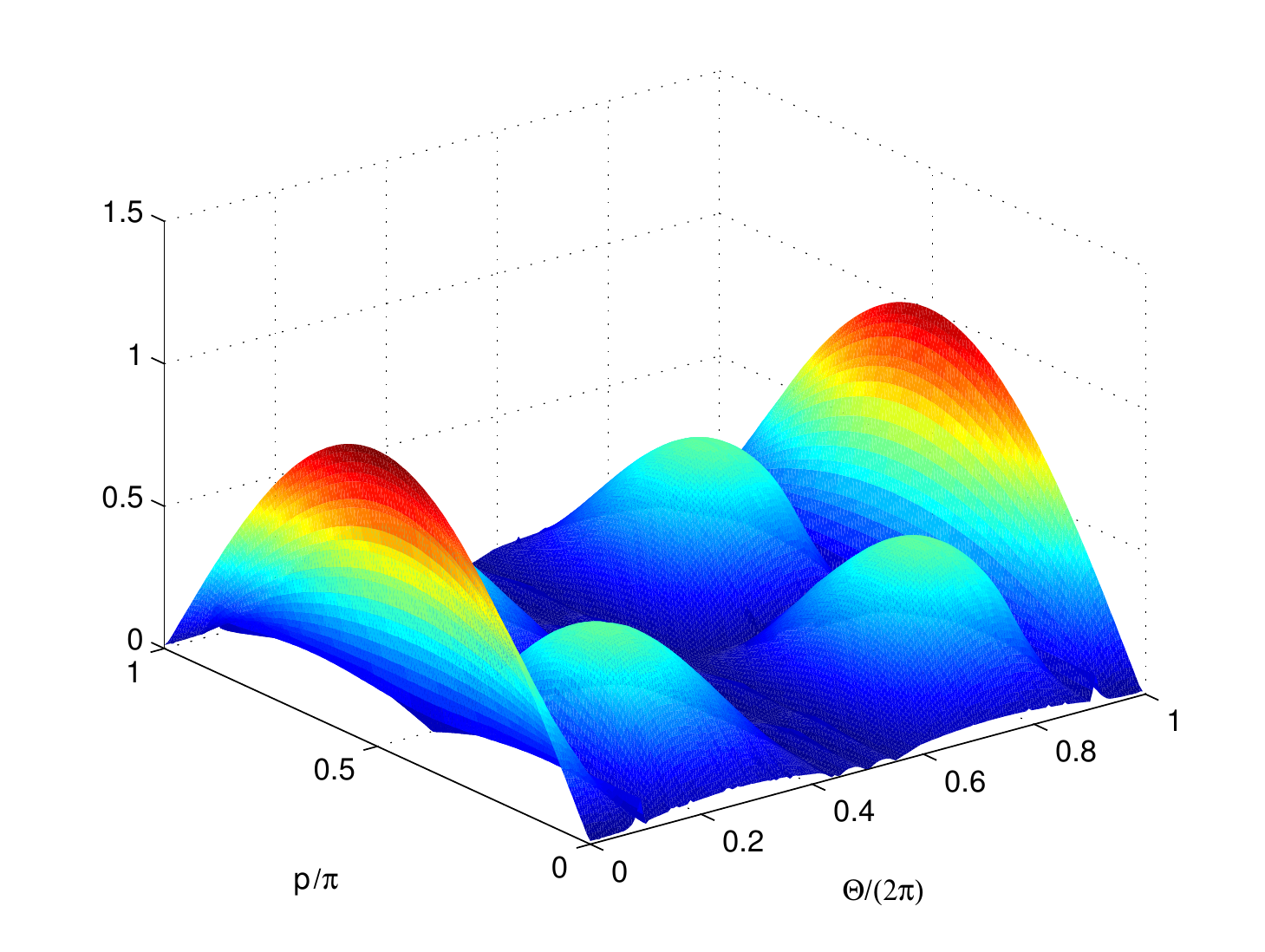}
\hspace*{\fill}
\includegraphics[width=0.45\textwidth]{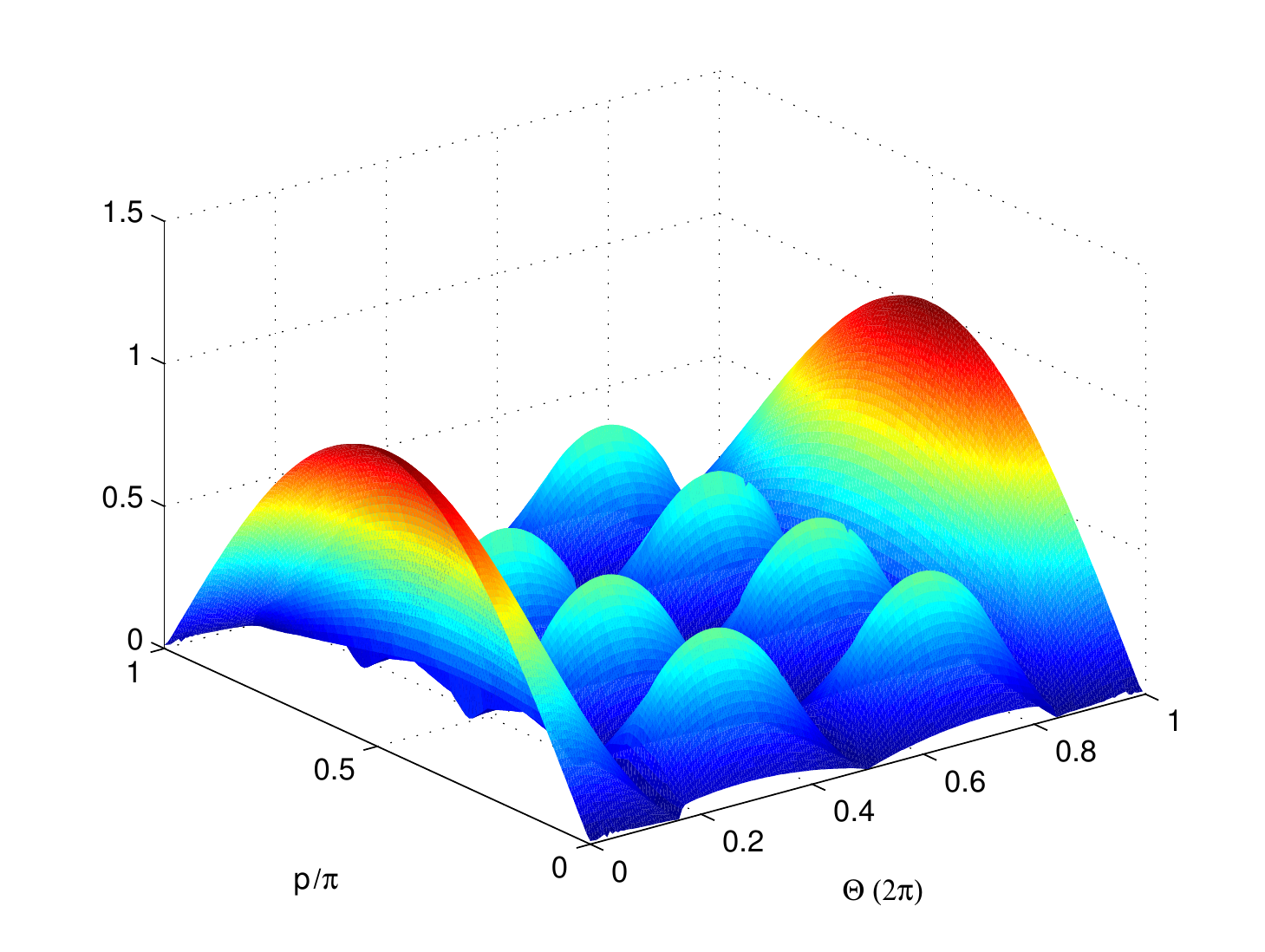}

 \begin{tikzpicture}[scale=2,
            point/.style={draw,scale=0.7,thick,fill=white},
            trans/.style={draw,scale=0.7,thick,fill=white,circle},
            labell/.style={inner sep=7pt,left},
            labelr/.style={inner sep=7pt,right}
        ]

        \draw[thick] (0,0) -- (0:1) arc (0:180:1) -- cycle;
        \draw[thick] (0,0) -- (180:1) arc (180:360:1) -- cycle;
        \draw[thick,fill=blue!20] (0,0) -- (36:1) arc (36:162:1) -- cycle;
        \draw[thick,fill=blue!20] (0,0) -- (198:1) arc (198:324:1) -- cycle;
        \draw[thick,fill=green!20] (0,0) -- (-2:1) arc (-2:2:1) -- cycle;
        \node at (120:0.6) {$c=\frac12+1$};
        \node [trans] at (0,0) {};
        \node [point] at (0:1) {};
        \node [point] at (62.2:1) {};
        \node [point] at (297.8:1) {};
        \node [trans] at (36:1) {};
        \node [trans] at (162:1) {};
        \node [point] at (117.8:1) {};
        \node [point] at (242.2:1) {};
        \node [trans] at (180:1) {};
        \node [trans] at (324:1) {};
        \node [trans] at (198:1) {};
        \node[inner sep=1.5pt] (a) at (345:0.70) {gapped};
        \draw (0.7,-0.10) -- (0.9,0.00);
        \node [labelr] at (0:1) {$\theta=0$};
        \node [labelr] at (62.2:1.05) {$(5,3,+1)$}; 
        \node [labell] at (117.8:1.05) {$(5,3,-1)$}; 
        \node [labell] at (180:1) {$\pi$};
        \node [labell] at (242.2:1.05) {$(5,3,-1)$}; 
        \node [labelr] at (297.8:1.05) {$(5,3,+1)$}; 

    \end{tikzpicture}
\hspace*{\fill}
   \begin{tikzpicture}[scale=2,
            point/.style={draw,scale=0.7,thick,fill=white},
            trans/.style={draw,scale=0.7,thick,fill=white,circle},
            labell/.style={inner sep=7pt,left},
            labelr/.style={inner sep=7pt,right}
        ]

        \draw[thick] (0,0) -- (0:1) arc (0:180:1) -- cycle;
        \draw[thick] (0,0) -- (180:1) arc (180:360:1) -- cycle;
        \draw[thick,fill=blue!20] (0,0) -- (60:1) arc (60:180:1) -- cycle;
        \draw[thick,fill=blue!20] (0,0) -- (180:1) arc (180:300:1) -- cycle;
        \draw[thick,fill=green!20] (0,0) -- (-2:1) arc (-2:2:1) -- cycle;
        \node at (120:0.6) {$c=1$};
        \node [trans] at (0,0) {};
        \node [point] at (0:1) {};
        \node [point] at (6:1) {};
        \node [trans] at (60:1) {};
        \node [point] at (174:1) {};
        \node [trans] at (180:1) {};
        \node [point] at (186:1) {};
        \node [trans] at (300:1) {};
        \node [point] at (354:1) {};
        \node[inner sep=1.5pt] (a) at (345:0.70) {gapped};
        \draw (0.7,-0.10) -- (0.9,0.00); 
        \node [labelr] at (0:1) {$\theta=0$};
        \node [labelr] at (6+5:1) {$(5,1,-1)$};
        \node [labell] at (174-5:1) {$(5,1,+1)$};
        \node [labell] at (180:1) {$\pi$};
        \node [labell] at (186+5:1) {$(5,1,+1)$};
        \node [labelr] at (354-5:1) {$(5,1,-1)$};


    \end{tikzpicture}

    \caption{Ground state expectation values of the local projection
      operators (top row) and dispersion of lowest excitations (middle
      row) of the $so(5)_2$ anyon chain as a function of the coupling
      constant $\theta$ for $\ell=1$ (left column) and $\ell=3$ (right
      column). The symbols $\blacktriangle$ mark the position of the
      FZ integrable points, red dashed lines indicate the conjectured
      locations of phase transitions at $\theta\simeq\pi/5$, $9\pi/10$
      for $\ell=1$ ($\theta\simeq\pi/3$ for $\ell=3$).  The bottom row
      shows the proposed phase diagram for the $SO(5)_2$ anyon chains.
      The Bethe ansatz results indicate that there is a gapped region
      surrounding the Temperley-Lieb integrable point ($\theta=0$).
      \label{fig:phasp_n5}}
\end{figure}
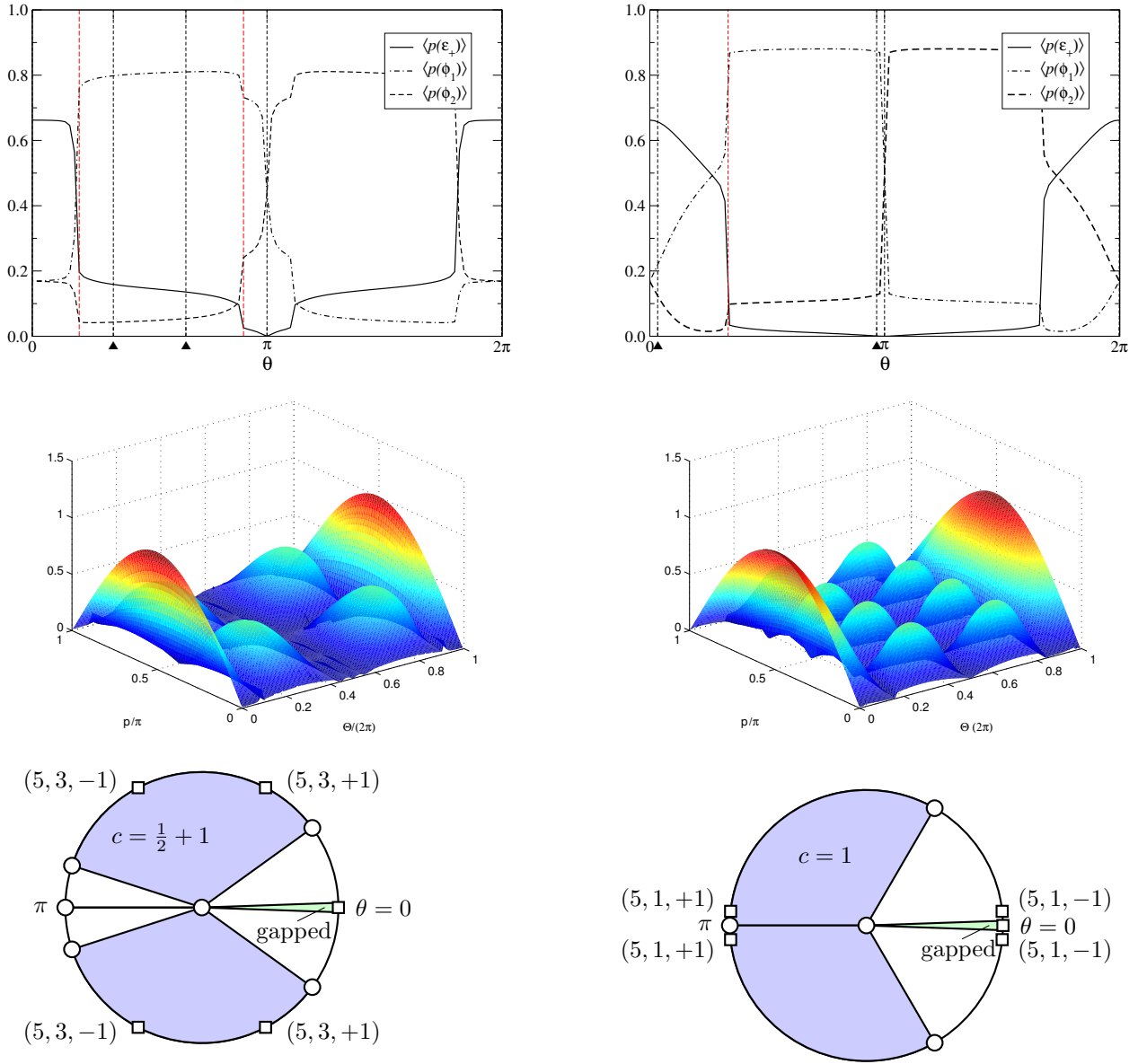

When $\theta$ is varied, the models undergo a sequence of phase transitions.
To identify the nature of the different phases we have used the
\textsc{evomps} software package \cite{evomps} to compute variational matrix
product states representing the ground state and the lowest excitation for a
given momentum.  In Fig.~\ref{fig:phasp_n5} the ground state expectation
values of the local projection operators and the dispersion for the $so(5)_2$
anyon models are shown.
From the numerical data the $\mathbf{Z}_2$ symmetry of the model under $\theta
\leftrightarrow 2\pi-\theta$ with a simultaneous exchange of $\phi_1
\leftrightarrow \phi_2$ is clearly seen.  

In the data for the model with $\ell=1$ we observe a singular change of the
expectation values and a change in the translational properties of the ground
state reflected in the periodicity of the dispersion of lowest excitations at
$\theta\simeq\pi/5$ and $9\pi/10$ which indicates the presence of phase
transitions.  The two FZ integrable points of $\mathcal{H}_{(5,\ell=1)}$ at
$\theta=\eta'$ and $\pi-\eta'$ with
$\eta'=\pi/4-\arctan\left(\frac{1-\sqrt{5}}{4}\right)$ are located between
these transition points.  We shall analyze the spectrum at these points in
more detail below.

The phase portrait of $\mathcal{H}_{(5,\ell=3)}$ has already been considered
previously: a phase transition where the expectation values of the projection
operators change in a singular way and the periodicity of dispersion of low
energy excitation switches to a different value is observed at $\pi\simeq
\pi/3$ \cite{Finch.etal14}.  

The low energy effective theories at the $\ell=3$ FZ integrable points have
been identified with rational conformal field theories respecting the
five-fold discrete symmetries of the anyon model which are invariant under
extensions of the Virasoro algebra \cite{FiFF14}, i.e.\ from the minimal
series of Casimir-type $\mathcal{W}$-algebras associated with the Lie-algebras
$B_2=SO(5)$, $D_5=SO(10)$, and $\mathcal{B}_{0,2}=OSp(1|4)$.

\paragraph*{\underline{$(n,\ell_{FZ},J)=(5,1,-1)$.}}
The low energy effective theory for the feromagnetic $\mathbf{Z}_5$ clock model is the $\mathbf{Z}_5$ parafermion CFT \cite{ZaFa85,JiMO86}.  In the finite size spectrum of the corresponding model of $so(5)_2$ anyons (i.e.\ $\theta=\eta$ with $\eta \equiv \pi/4-\arctan\left(\frac{1+\sqrt{5}}{4}\right)$) additional conformal weights appear implying that the continuum limit is described by a $\mathcal{W}B_2(5,7)$ rational CFT with the same central charge $c=\frac87$ and conformal weights (\ref{cft_specWB2-57}), see \cite{FiFF14}.

\paragraph*{\underline{$(n,\ell_{FZ},J)=(5,1,+1)$.}}
Similarly, the continuum limit of the \emph{anti}ferromagnetic integrable
model $(n,\ell_{FZ},J)=(5,1,+1)$ corresponding to $\theta=\pi-\eta$ has been
found to be described by a $\mathcal{W}D_5(9,10)$ rational CFT with $c=1$ and conformal weights (\ref{cft_specWD5-9-10}), equivalent to the $\mathbf{Z}_2$-orbifold of a Gaussian model with compactification radius $2R^2=5$.

For the critical properties of the FZ integrable points
$(n,\ell_{FZ},J)=(5,3,\pm1)$ we have analyzed the Bethe equations based on the
root density formalism (see Appendix~\ref{app:thermo}).  This approach
provides the ground state energy density and Fermi velocities of low energy
excitations in the thermodynamic limit.  Based on these data the central
charge and scaling dimensions can be extracted from the finite size spectra
(\ref{fscft}).

\paragraph*{\underline{$(n,\ell_{FZ},J)=(5,3,+1)$}.} 
From the analysis of the thermodynamic limit the Bethe root configuration for the ground state is known to consist of $3L/2$ $(2,+)$-strings and $L/2$ $(2,-)$ strings, see Table~\ref{table:thermo357}.  The ground state energy density is $\epsilon_\infty= -6.61811809391087$ and the Fermi velocity of low lying excitations is $v^{(F)}=5/2$.
The finite size scaling analysis (\ref{fscft}) of the spectrum of this model using data obtained from the solution of the Bethe ansatz gives a central charge $c=3/2$ and conformal weights as listed in
Table~\ref{table:fsdata53p}.
\begin{table}[t] 
\begin{ruledtabular}
\begin{tabular}{ccccccccc} 
  $X$ (extrap.) & $s$ & $(h,\bar{h})_1 + (h,\bar{h})_2$ & $\Delta_{(1,+)}$ & $\Delta_{(1,-)}$ & $\Delta_{(2,+)}$ & $\Delta_{(2,-)}$ & $\Delta_{(1,m)}$ & comment \tabularnewline \hline
  0.125000 & 0 & $(0,0) + (\frac{1}{16},\frac{1}{16})$ & 0 & $2$ & 0 & $-1$ & 0 & a \tabularnewline 
  0.175000 & 0 & $(\frac{1}{16},\frac{1}{16}) + (\frac{1}{40},\frac{1}{40})$ 
    & 1 & 1 & $-1$ & $-1$ & 0 & a,c \tabularnewline 
  0.200000 & 0 & $(0,0) + (\frac{1}{10},\frac{1}{10})$ & 0 & 0 & $-1$ & $-1$ & 0 & a,c \tabularnewline 
  0.250000 & 0 & $(\frac{1}{16},\frac{1}{16})+(\frac{1}{16},\frac{1}{16})$ 
    & 0 & 0 & 0 & 0 & 0 & a \tabularnewline 
  0.575000 & 0 & $(\frac{1}{16},\frac{1}{16}) + (\frac{9}{40},\frac{9}{40})$ 
    & 1 & 3 & $-2$ & $-2$ & 0 & a,c \tabularnewline 
  0.800000 & 0 & $(0,0) + (\frac{2}{5},\frac{2}{5})$ & 2 & 0 & $-2$ & 0 & 0 & a,c \tabularnewline 
  1.000000 & 1 & descendant & 1 & 1 & $-1$ & $-1$ & 1 & a,c \tabularnewline 
  1.000000 & 0 & $(\frac{1}{2},\frac{1}{2}) + (0,0)$ & 0 & 0 & $-1$ & $-1$ & 2 & a,c \tabularnewline 
  1.125000 & 0 & $(0,0) + (\frac{9}{16},\frac{9}{16})$ & 0 & 2 & $-1$ & $-2$ & 2 & a \tabularnewline 
\end{tabular}
\end{ruledtabular}
\caption{\label{table:fsdata53p}
  Finite size data for the effective scaling dimensions from the solution of   the Bethe equations for the low lying excitations of the $(n,\ell_{FZ},J)=(5,3,+1)$ model.  The conjectured weights $(h,\bar{h})_1+(h,\bar{h})_2$ are based on our proposal of a low energy theory with two critical factors, namely an Ising model ($c_1=\frac12$) and a rational CFT with central charge $c_2=1$. In the last column we indicate in which model (a: anyon chain, c: clock model) the corresponding level can be observed -- possibly subject to further constraints on the system size $L$. 
  In addition to the levels listed there is a spin $\frac{1}{2}$ state in the anyon chain with scaling dimension $\frac{23}{40}<X=\sum h_i+\bar{h}_i <\frac{4}{5}$. Unfortunately we have not been able to identify its Bethe root configuration and have no numerical data for sufficiently large system sizes allowing for a finite size extrapolation.  Based on our proposal for the underlying CFT we conjecture that this missing level corresponds to the product of the identity in the Ising sector and a primary field with conformal weights $(h,\bar{h})_2 = (\frac{9}{16},\frac{1}{16})$ in sector $2$.   
  }  
\end{table}
Based on the observed spectrum we conjecture that the critical theory is a product of an Ising model ($c_1=\frac12$) and a rational CFT with central charge $c_2=1$.  We have found conformal weights $h\in\{0,\frac{1}{40}, \frac{1}{16}, \frac{1}{10}, \frac{9}{40}, \frac25\}$, which is consistent with the $\mathbf{Z}_2$-orbifold of a $U(1)$ boson compactified with radius $2R^2=10$ (coinciding with the $\mathcal{W}$-minimal model $\mathcal{W}D_{10}(19,20)$, see (\ref{cft_specWD10})), or the $\mathcal{WB}_{0,2}(4,5)$ minimal model with conformal weights (\ref{cft_specWBf2-45}). 
Note that these data alone are not sufficient to distinguish between these two rational CFTs. Even if we could compute data for states with larger conformal weights, contained in the spectrum of the $\mathcal{W}D_{10}(19,20)$ algebra, but not in the spectrum of the $\mathcal{WB}_{0,2}(4,5)$ algebra, we could not make a choice. The reason is that the additional conformal weights $h'$ in the spectrum of $\mathcal{W}D_{10}(19,20)$ all differ by non negative half-integers or integers from the conformal weights $h$ in $\mathcal{WB}_{0,2}(4,5)$,
$h'=h+k/2$ with $k\in\mathbb{Z}_+$, see Appendix~\ref{app:rCFTs-WBf}. This happens because the extended chiral symmetry algebra of the $\mathcal{WB}_{0,2}(4,5)$ model possesses a generator $Q$ of half-integer scaling dimension $h_Q=5/2$. This generator allows to build states from representations with conformal dimensions $h'$ shifted by an half-integer or integer as excitations by modes of this generator of states from representations with unshifted conformal weights $h$, e.g.\ $|h'\rangle = Q_{-k/2}|h\rangle$. Thus, we expect that the $\mathcal{W}D_{10}(19,20)$ model admits a non-diagonal partition function with terms of a form like $|\chi^{D_{10}}_h+\chi^{D_{10}}_{h'}|^2$, as linear combinations of the corresponding characters yield the characters of the representations of the $\mathcal{WB}_{0,2}(4,5)$ model, which would then read something like $\chi^{D_{10}}_h+\chi^{D_{10}}_{h'}=\chi^{\mathcal{B}_{0,2}}_h$.

A hint towards the identification can be taken from the degeneracies in the spectrum found from the exact diagonalization of the lattice models of lengths up to $L=10$: in the clock model we find that all levels apart from those containing the singlet vacuum $(h,\bar{h})_2=0$ of the $c_2=1$ sector appear with even degeneracy.  This agrees with what is found in the $\mathcal{WB}_{0,2}$ model as a consequence of existence of the fermionic field $Q$ with half-integer modes, see Appendix~\ref{app:rCFTs-WBf}.  The multiplicities in the spectrum of the corresponding anyon chain, however, are different: among the spin $s=0$ levels listed in Table~\ref{table:fsdata53p} only the one with weights $(\frac{1}{16},\frac{1}{16}) +(\frac{1}{16},\frac{1}{16})$ appears with multiplicity two for the system sizes which can be handled by numerical diagonalization.

In order to really identify the rational CFT, one would have to identify all good quantum numbers of the lattice models and map them to the weights of the representations with respect to the zero modes of all generators of the respective chiral symmetry algebras. If one such quantum number is to be identified with the eigenvalues of the zero mode $W_0$ of the generator with half-integer scaling dimension, the choice would necessarily be the $\mathcal{WB}_{0,2}(4,5)$ model. So far, the numerical data only gives us the weights with respect to the Virasoro generator $L_0$, which is associated to the energy quantum number.

\paragraph*{\underline{$(n,\ell_{FZ},J)=(5,3,-1)$}}
The thermodynamic Bethe ansatz analysis predicts a ground state energy density
$\epsilon_\infty=-2.75801611466$ and that this model has two branches of low
energy excitations with different Fermi velocities, $v^{(F)}_{1}=5$ and
$v^{(F)}_{2}=5/3$, respectively (see Table~\ref{table:thermo357}).  This
indicates that the effective field theory is a product of two sectors.
Unfortunately, the numerical solution of the Bethe equations is plagued by
instabilities and we have to rely on data obtained using the Lanczos algorithm
for systems sizes of up to $L=11$.  Extrapolating the finite size data of the
ground state energy (realized for even $L$ in the clock model and for $L=0\mod
8$ for the anyon chain) we find
\begin{equation}
  \label{C2-extra}
  C(L) \equiv -\frac{6L}{\pi} \left(E_0(L) -L\epsilon_\infty\right)
    \to \frac{25}{6}\,,
\end{equation}
see Table~\ref{table:fsdata53m}.  For a CFT with two critical degrees of freedom this quantity is expected to be the combination $\left( v^{(F)}_{1}\, c_1  + v^{(F)}_{2}\, c_2\right)$ of the Fermi velocities and the universal central charges $c_i$ of the factor CFTs.  Assuming that both factors are unitary the unique solution is $c_1=\frac12$, $c_2 = 1$, implying that the continuum limit of the lattice models is described by an Ising model and a $c=1$ CFT.
\begin{table}[t]
\begin{ruledtabular}
\begin{tabular}{c|ccccc|c|c|c}
    $L$ & $2$ & $4$ & $6$ & $8$ & $10$ & extr.\ & conj.\ & \\
    \hline
    $C(L)$ &    4.5466656 
            &   4.2580757 
            &   4.2070680 
            &   4.1893536 
            &   4.1811756 
    & $4.166(1)$ & $\frac12+1$ & a,c
  \\\hline 
  $X(L)$ &  0.1982213 
         &  0.2060728 
         &  0.2073608 
         &  0.2077947 
         &  0.2079917 
    & 0.2083(2) &  $(0,0)+(\frac{1}{16},\frac{1}{16})$ & a
  \\ 
        &  0.3306743 
          &  0.3328838 
          &  0.3331548 
          &  0.3332373 
          &  0.3332732 
   & 0.3333(3) & $(0,0)+(\frac{1}{10},\frac{1}{10})$ & a,c
 \\ 
     &   0.7237420 
       &  0.7120260 
       &  0.7099619 
       &  0.7092469 
       &  0.7089172 
   & 0.7083(1) & $(\frac{1}{16},\frac{1}{16})+(\frac{1}{40},\frac{1}{40})$ & a,c
 \\ 
     &  0.8534244 
      & 0.8380366 
      & 0.8354018 
      & 0.8344928 
      & 0.8340742 
   & 0.8333(1) & $(\frac{1}{16},\frac{1}{16})+(\frac{1}{16},\frac{1}{16})$ & a
 \\ 
      &   1.5861812 
      &   1.3565224 
      &   1.3430229 
      &   1.3386661 
      &   1.3367116 
  & 1.333(1) & $(0,0)+(\frac25,\frac25)$ & a,c
 \\ 
    &  1.3868122 
    &  1.3784210 
    &  1.3765362 
    &  1.3758661 
    &  1.3755547 
  & 1.3749(3) & $(\frac{1}{16},\frac{1}{16})+(\frac{9}{40},\frac{9}{40})$ & a,c
 \\\hline
    $L$ & $3$ & $5$ & $7$ & $9$ & $11$ &  &  &\\
    \hline
  $X(L)$ &    0.7150004 
           &  0.7106879 
           &  0.7095284 
           &  0.7090548 
           &  0.7088157 
  & 0.7083(1) & $(\frac{1}{16},\frac{1}{16})+(\frac{1}{40},\frac{1}{40})$ & a,c
 \\ 
    &  0.7670079 
    & 0.7920987 
    & 0.8033895 
    & 0.8098203 
    & 0.8139754 
  & 0.8332(2) & $(\frac{1}{16},\frac{1}{16})+(\frac{1}{16},\frac{1}{16})$ & a
 \\ 
    &   0.9160522 
      &  0.8804778 
      &  0.8662928 
      &  0.8586707 
      &  0.8539125 
   & 0.833(1) & $(\frac{1}{16},\frac{1}{16})+(\frac{1}{16},\frac{1}{16})$ & a
 \\ 
    &   0.9748811 
     &  1.0175706 
     &  1.0294044 
     &  1.0342663 
     &  1.0367220 
     & 1.041(2) & $(0,0)+(\frac{9}{16},\frac{1}{16})$ & a,2
 \\ 
    &   1.3813560 
    &   1.3772372 
    &   1.3761362 
    &   1.3756863 
    &   1.3754591 
    & 1.375(1) & $(\frac{1}{16},\frac{1}{16})+(\frac{9}{40},\frac{9}{40})$ & a,c
\end{tabular}
\end{ruledtabular}

\caption{\label{table:fsdata53m}
  Numerical finite size data obtained using the Lanczos algorithm for the finite size scaling of the ground state energy  (\ref{C2-extra}) and the scaled energy gaps (\ref{X2-extra}) of the $(n,\ell_{FZ},J)=(5,3,-)$ model for small systems together with their VBS extrapolation \cite{HaBa81}. The conjectures are central charges $c_1+c_2$ and conformal weights $(h,\bar{h})_1+(h,\bar{h})_2$ based on the proposed factorization into an Ising model and a Gaussian model with $\mathbf{Z}_2$ orbifold compactification radius $2R^2=10$ for the continuum limit. 
  In the last column we indicate in which model (a: anyon chain, c: clock model) the corresponding level can be observed -- possibly subject to further constraints on the system size $L$.  Similarly, a label $2$ indicates a spin $\frac12$ state: this has been used in the identification of the state with conformal weights $(0,0)+(\frac{9}{16},\frac{1}{16})$.}  
\end{table}
Similarly, we extrapolate the quantities
\begin{equation}
  \label{X2-extra}
  X_n(L) \equiv \frac{L}{2\pi} \left(E_n(L) - E_0(L)\right)\,
\end{equation}
for the low lying excitations to identify the operator content of the two sectors. In the thermodynamic limit $X(L) \to \left( v^{(F)}_{1}\, X_{1,n} + v^{(F)}_{2}\, X_{2,n}\right)$ where $X_{i,n} = h_{i,n} + \bar{h}_{i,n}$ is the scaling dimension of an operator with conformal weights $(h_{i,n},\bar{h}_{i,n})$ in sector $i=1,2$.
Our numerical data indicate that the spectrum of conformal weights in the $c=1$ sector coincide with those observed in the $(5,3,+1)$ model discussed above.  As a consequence, candidates for the CFT in this sector are again the Gaussian model with $\mathbf{Z}_2$-orbifold compactification with radius $2R^2=10$ (or, equivalently, one of  the $\mathcal{W}D_{10}(19,20)$ or $\mathcal{WB}_{0,2}(4,5)$ minimal models).
Together with the numerical data for ground state phase diagram of the generic Hamiltonian $\mathcal{H}_{(5,1)}$ (\ref{hamil5}) discussed above this leads us to conjecture an extended phase of this model with an effective central charge $c=\frac12+1$ for coupling constants $\pi/5<\theta<9\pi/10$, see Fig.~\ref{fig:phasp_n5}~(a).  Within this phase the Fermi velocities are non-universal functions of the parameter $\theta$.

%
\subsection{The $\mathbf{Z}_{7}$ and $so(7)_{2}$ models}
\begin{figure}[t]
\includegraphics[width=0.84\textwidth]{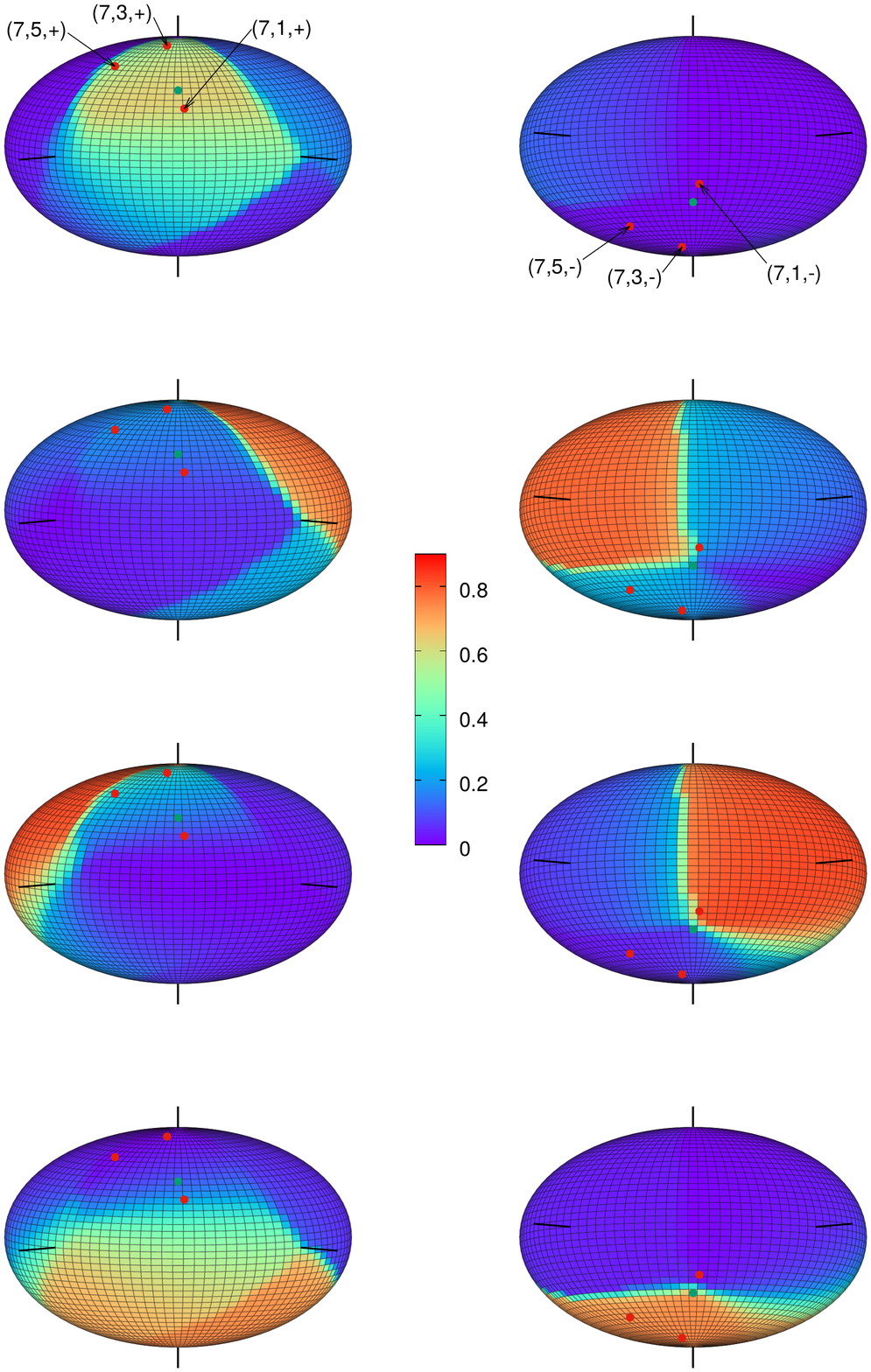}
\caption{Ground state expectation values of the local projection operators ($\langle p^{(\epsilon_+)} \rangle$, $\langle p^{(\phi_1)} \rangle$, $\langle p^{(\phi_2)} \rangle$, $\langle p^{(\phi_3)} \rangle$ from top to bottom) of the $so(7)_2$ anyon chain as a function of the coupling constants.  Data are colorcoded on the spherical parameter space of of the models (\ref{hamil7_sphere}) in terms of the projectors obtained from the $\ell=1$ $F$-moves.  Red dots indicate the location of the Fateev-Zamolodchikov integrable points $(7,\ell_{FZ},J)$, similarly the Temperley-Lieb integrable models are located at the green dots.
  \label{fig:phasp_n7}}
\end{figure}

As discussed above, the Hamiltonians $\mathcal{H}_{(7,\ell)}$ of the clock and
anyon models are equivalent under the unitary transformations constructed from
the automorphisms $\nu_t$ (\ref{map-nut}) and (\ref{maps-son}), respectively.
Therefore the integrable points correspond to particular choices of the
coupling constants $\boldsymbol{\alpha}$ on the surface of a $3$-sphere.  For
the presentation of the phase portrait it is convenient to choose two angular
variables to parametrize the coupling constants, i.e.
\begin{equation}
  \label{hamil7_sphere}
  \mathcal{H}_{(7)}
  (\theta,\psi) = \sum_{j=1} \left[
    \cos(\psi) \sin(\theta) \, p_j^{(\phi_{1})} 
    + \sin(\psi) \sin(\theta) \, p_j^{(\phi_{2})}
    + \cos(\theta) \, p_j^{(\phi_{3})} \right]\,.
\end{equation}
In this parameterization the Temperley--Lieb points occur at
$(\psi_{TL},\theta_{TL}) = (\frac{\pi}{4}, \arctan(\sqrt{2}))$ and
$(\psi_{TL}+\pi,\pi-\theta_{TL})$.  To illustrate some properties of the
phases of the model as the coupling parameters are varied ground state
expectation values $\langle p^{(\epsilon_+)} \rangle$, $\langle p^{(\phi_1)}
\rangle$, $\langle p^{(\phi_2)} \rangle$, $\langle p^{(\phi_3)} \rangle$ as
obtained from the variational matrix product ground states computed
numerically using the \textsc{evomps} library \cite{evomps} are shown in
Fig.~\ref{fig:phasp_n7}.
Note that, as in the $n=5$ model, the ferromagnetic TL point (located on the southern hemisphere) is located at the intersection of $(n-1)/2$ phases differing in the label $k$ of the dominant order parameter $\langle p^{(\phi_k)} \rangle$ leading to a highly degenerate spectrum.  Similarly, the model is expected to be gapped in a neighbourhood of the \emph{anti}ferromagnetic TL point (on the northern hemisphere).

For the FZ integrable points we have used the approach described above to
compute the central charge(s) and some of the scaling dimensions of primary
fields of the low energy effective theories describing the continuum limit of the lattice models.


\paragraph*{\underline{$(n,\ell_{FZ},J)=(7,1,+1)$}.}
The configuration of Bethe roots corresponding to the ground state of this model consists of $(1,\pm)$-strings, see Table~\ref{table:thermo357}.  From the root density approach we obtain the ground state energy density, $\epsilon_\infty = -11.7779598163070$.  There is a single branch of gapless low energy excitations with Fermi velocity $v^{(F)}=7/6$.  From the scaling analysis (\ref{fscft}) of the finite size spectra obtained from the solution of the Bethe equations we obtain a central charge $c=1$ and the conformal weights shown in Table~\ref{table:fsdata71p}. 
These data are consistent with the $\mathbb{Z}_2$-orbifold of a Gaussian model with compactification radius $2R^2=7$, see Eq.~(\ref{cft_specWD7-1314}),  and in agreement with the conjectured $\mathcal{W}D_n(2n-1,2n)$ rational CFT description of the critical series of \emph{anti}ferromagnetic $\mathbf{Z}_n$ FZ models
with $c=1$ \cite{FiFF14}.
\begin{table}[t] 
\begin{ruledtabular}
\begin{tabular}{lccccc}
  $X$ (extrap.) & $s$ & $(h,\bar{h})$ & $\Delta_{(1,+)}$ & $\Delta_{(1,-)}$ & comment \tabularnewline \hline
  0.071429 & 0 & $(\frac{1}{28},\frac{1}{28})$ & $-1$ & $-1$ & a,c \tabularnewline 
  0.125000 & 0 & $(\frac{1}{16},\frac{1}{16})$ & 0 & 0 & a \tabularnewline 
  0.285714 & 0 & $(\frac{1}{7},\frac{1}{7})$ & $-2$ & $-2$ & a,c \tabularnewline 
  0.642857 & 0 & $(\frac{9}{28},\frac{9}{28})$ & $-3$ & $-3$ & a,c \tabularnewline 
  1.142856(5) & 0 & $(\frac{4}{7},\frac{4}{7})$ & $-4$ & $-2$ & a,c \tabularnewline 
\end{tabular}
\end{ruledtabular}
\caption{\label{table:fsdata71p}
  Similar as Table~\ref{table:fsdata53p} but for the $(7,1,+1)$ model.  The conjectured conformal weights are from a $\mathcal{W}D_7(13,14)$ minimal model (\ref{cft_specWD7-1314}).}
\end{table}


\paragraph*{\underline{$(n,\ell_{FZ},J)=(7,1,-1)$}.}
From the Bethe ansatz analysis of the thermodynamic limit we obtain the ground state energy density $\epsilon_\infty = -4.34504525693274$ of this model as well as the Fermi velocity $v^{(F)}=7$ of the single branch of its gapless excitations.
This ferromagnetic FZ clock model has been identified with $\mathbf{Z}_7$ parafermion CFT \cite{ZaFa85,JiMO86} with central charge $c=\frac43$ and conformal weights (\ref{cft_specZ7}).  Indeed the lowest $\mathbf{Z}_7$ conformal weights are found in the finite size spectrum of the clock model, see Table~\ref{table:fsdata71m}.  In the spectrum of the corresponding $so(7)_2$ anyon chain, however, additional levels are present with $(h,\bar{h}) = (\frac{1}{24}, \frac{1}{24})$ and $(\frac{7}{72},\frac{7}{72})$ leading to conjecture that the critical theory is the $\mathcal{W}B_3(7,9)$ rational CFT which has the same central charge and conformal weights listed in (\ref{cft_specWB3-79}).  As noted in Appendix~\ref{app:rCFTs} the spectrum of $\mathbf{Z}_7$ is a subset hereof. The observed anyon levels are the $\mathcal{W}B_3(7,9)$ primaries with the lowest weights outside the parafermion subset
\begin{equation}
\label{cft_specWB3-79x}
  \mathrm{spec}(\mathcal{W}B_3(7,9)) \setminus \mathrm{spec}(\mathbf{Z}_7)
      = \left\{\frac{1}{24}, \frac{7}{72}, \frac{5}{24}, 
           \frac38, \frac{13}{24}, \frac{43}{72},
           \frac{17}{24}, \frac{11}{9}, \frac53, \frac{15}{8}, 
           \frac73, 3 \right\}\,.
\end{equation}
\begin{table}[t] 
\begin{ruledtabular}
\begin{tabular}{lccc}
  $X$ (extrap.) & $s$ & $(h,\bar{h})$ & comment \tabularnewline \hline
  0.08345(3) & 0 & $(\frac{1}{24},\frac{1}{24})$ & a \tabularnewline  
  0.09612 & 0 & $(\frac{1}{21},\frac{1}{21})$ & a,c \tabularnewline 
  0.159001 & 0 & $(\frac{5}{63},\frac{5}{63})$ & a,c \tabularnewline 
  0.190512 & 0 & $(\frac{2}{21},\frac{2}{21})$ & a,c \tabularnewline 
  0.195(1) & 0 & $(\frac{7}{72},\frac{7}{72})$ & a,* \tabularnewline  
\end{tabular}
\end{ruledtabular}
\caption{\label{table:fsdata71m}
  Similar as Table~\ref{table:fsdata53p} but for the $(7,1,-1)$ model.  The conjectured conformal weights appear in the $\mathcal{W}B_3(7,9)$ rational CFT.  We do not list Bethe root configurations $\Delta_\gamma$ here, as the root patterns identified in Appendix~\ref{app:thermo} are obscured by finite size effects.  For the level (*) extrapolating to $h=\bar{h}=\frac{7}{72}$ we have no Bethe ansatz results and use Lanczos data instead.}
\end{table}

\paragraph*{\underline{$(n,\ell_{FZ},J)=(7,3,+1)$}.}
From the Bethe ansatz analysis of the thermodynamic limit we obtain the ground state energy density of this model to be $\epsilon_\infty = -6.00247204898418$.  There are three branches of low lying excitations over the ground state with two different Fermi velocities, $v^{(F)}_1=7$ and $v^{(F)}_2=7/4$.  We have not succeeded in solving the Bethe equations for finite chains.  Instead we rely on numerical finite size data obtained using the Lanczos algorithm for the identification of the critical properties of this model: based on our extrapolation of the numerical data for $C(L)$ as defined in (\ref{C2-extra}) we conjecture the critical theory to be a product of two sectors with $c_1=4/5$ and $c_2=1$, respectively, giving $C(L)\to 147/20$, see Table~\ref{table:fsdata73p}.
\begin{table}[t] 
\begin{ruledtabular}
\begin{tabular}{c|ccccc|c|c|c}
    $L$ & $2$ & $4$ & $6$ & $8$ & $10$ & extrap. & conj. &\\
    \hline
    $C(L)$ &   $8.0776124$ & $7.5286196$ & $7.4302729$ & $7.3956872$ & $7.3795512$
    & $7.35(1)$ & $\frac45+1$ & a,c\\\hline
    $X(L)$ &  0.3509154 
           &  0.3693661 
           &  0.3724672 
           &  0.3735447 
           &  0.3740478 
         & 0.375(2) & $(0,0)+(\frac{3}{28},\frac{3}{28})$ & a,c\\
         &  -- & 0.5765752 
           &   -- & 0.5735123 
           &  -- 
           & -- & $(\frac{1}{40},\frac{1}{40})+(\frac{1}{16},\frac{1}{16})$ & a\\
         &  1.0158173 
           &  0.9931704 
           &  0.9871721 
           &  0.9843406 
           &  0.9826656 
         & 0.974(3)& $(\frac{1}{15},\frac{1}{15})+(\frac{1}{84},\frac{1}{84})$ & a,c\\
         &   1.1435201 
           &   1.1190081 
           &   1.1126025 
           &   1.1096049 
           &   1.1078447 
         & 1.100(3) & $(\frac{1}{15},\frac{1}{15})+(\frac{1}{21},\frac{1}{21})$ & a,c \\
         &   1.6262976 
           &   1.5218150 
           &   1.5085575 
           &   1.5045289 
           &   1.5045289 
         & 1.500(2) & $(0,0)+(\frac37,\frac{3}{7})$ & a,c\\
         &   1.9150465 
           &   1.6157237 
           &   1.6112404 
           &   1.6088639 
           &   1.6074000 
         & 1.604(5) & $(\frac{1}{15},\frac{1}{15})+(\frac{4}{21},\frac{4}{21})$ & a,c\\
         &   ---
           &   1.5564081 
           &   1.6636580 
           &   1.7013413 
           &   1.7188205 
         & 1.73(2) & $(\frac{1}{15},\frac{1}{15})+(\frac{25}{84},\frac{25}{84})$ & c\\
\end{tabular}
\end{ruledtabular}
\caption{\label{table:fsdata73p}
  Similar as Table~\ref{table:fsdata53m} but for the $(7,3,+1)$ model.  The last column contains our conjectures for $c_1+c_2$ and $(h,\bar{h})_1+(h,\bar{h})_2$ based on a factorization of the low energy effective theory with the first factor being the $\mathcal{M}_{(5,6)}$ minimal model with central charge $c_1=\frac45$ and conformal weights (\ref{cft_specPotts3}).}
\end{table}
Identifying the first factor with the $\mathcal{M}_{(5,6)}$ minimal model we also obtain conjectures for some of the lowest scaling dimensions present in the $c_2=1$ sector. We observe that most of these are of the form $k^2/42$ with integer $k$.  In addition there is numerical evidence for a level with $X_2=\frac18$ in the spectrum of the anyon chain.  Based on this observation we conjecture the that the latter sector is a $\mathbf{Z}_2$-orbifold of a Gaussian model with compactification radius $2R^2=21$ with spectrum (\ref{cft_specWD21}), see also the finite size analysis for the $(7,5,+1)$ model below.

\paragraph*{\underline{$(n,\ell_{FZ},J)=(7,3,-1)$}.}
The energy density of this model in the thermodynamic limit is $\epsilon_\infty = -8.35336129420055$, gapless excitations are propagating with Fermi velocity $v^{(F)}= 7/3$.  From the finite size analysis based on the Bethe ansatz we find that the central charge of the critical theory is $c=3/2$.  We have also identified some conformal weights, see Table~\ref{table:fsdata73m}.
\begin{table}[t] 
\begin{ruledtabular}
\begin{tabular}{lcccccccc}
  $X$ (extrap.) & $s$ & $(h,\bar{h})_1+(h,\bar{h})_2$ & $\Delta_{(1,+)}$ & $\Delta_{(1,-)}$ & $\Delta_{(2,+)}$ & $\Delta_{(2,-)}$ & $\Delta_{(1,m)}$  &comment \tabularnewline \hline
  0.125000 & 0 & $(0,0)+(\frac{1}{16},\frac{1}{16})$ &&&&&& a,* \tabularnewline 
  0.142857 & 0 & $(0,0)+(\frac{1}{14},\frac{1}{14})$ & 0 & 0 & $-1$ & $-1$ & 0 & a,c \tabularnewline 
  0.160714 & 0 & $(\frac{1}{16},\frac{1}{16}) + (\frac{1}{56},\frac{1}{56})$
    & 0 & 0 & $-1$ & $-1$ & 1 & a,c \tabularnewline 
  0.250000 & 0 & $(\frac{1}{16},\frac{1}{16})+(\frac{1}{16},\frac{1}{16})$ & 2 & 0 & $-1$ & 0 & 0 & a \tabularnewline 
  &&& 0 & 2 & 0 & $-1$ & 0 & a \tabularnewline 
  0.446429 & 0 & $(\frac{1}{16},\frac{1}{16}) + (\frac{9}{56},\frac{9}{56})$
    & 0 & 0 & $-2$ & $-2$ & 1 & a,c \tabularnewline 
  0.571428(4) & 0 & $(0,0)+(\frac{2}{7},\frac{2}{7})$ & 0 & 2 & $-2$ & $-2$ & 0 & a,c \tabularnewline 
  1.000000 & 0 & $(\frac{1}{2},\frac{1}{2})+(0,0)$ & 2 & 2 & $-1$ & $-1$ & 0 & a,c
  \tabularnewline   
\end{tabular}
\end{ruledtabular}
\caption{\label{table:fsdata73m}
  Similar as Table~\ref{table:fsdata53p} but for the $(7,3,-1)$
  model.  We have not identified the Bethe root configuration for the anyon
  state with $(h,\bar{h}) = (\frac{1}{16},\frac{1}{16})$, the extrapolation is
  based on numerical finite size data from diagonalization of the Hamiltonian. For the level (*) extrapolating to $h_2=\bar{h}_2=\frac{1}{16}$ we have no Bethe ansatz results and use Lanczos data instead.} 
\end{table}
Based on these data we conjecture that the effective theory describing this model in the thermodynamic limit is again a product of two rational CFTs, namely an Ising model and a $\mathbb{Z}_2$-orbifold of a $U(1)$-boson compactified to a circle with radius $2R^2=14$.  The latter contributes $c_2=1$ and conformal weights (\ref{cft_specWD14}) to the finite size scaling.


\paragraph*{\underline{$(n,\ell_{FZ},J)=(7,5,+1)$}.}
The ground state energy density of this model is $\epsilon_\infty = -10.1731518217117$, its excitations are gapless and propagate with Fermi velocity $v^{(F)}=7/2$.  Analyzing the finite size spectrum obtained from the Bethe equations we find that the central charge is $c=9/5$ and have identified several conformal weights, see Table~\ref{table:fsdata75p}. 
\begin{table}[t] 
\begin{ruledtabular}
\begin{tabular}{lcccc}
  $X$ (extrap.)& $s$ & $(h,\bar{h})_1+(h,\bar{h})_2$  & comment \tabularnewline \hline
  0.157171 & 0 & $(\frac{1}{15},\frac{1}{15})+(\frac{1}{84},\frac{1}{84})$ & a,c \tabularnewline  
  0.175005(2) & 0 & $(\frac{1}{40},\frac{1}{40})+(\frac{1}{16},\frac{1}{16})$  & a \tabularnewline 
  0.214286 & 0 & $(0,0)+(\frac{3}{28},\frac{3}{28})$& a,c \tabularnewline 
  0.228576 & 0 & $(\frac{1}{15},\frac{1}{15})+(\frac{1}{21},\frac{1}{21})$& a,c \tabularnewline 
  0.375000 & 0 & $(\frac18,\frac18)+(\frac{1}{16},\frac{1}{16})$& a \tabularnewline 
  0.72864(2) & 0 & $(\frac{1}{15},\frac{1}{15})+(\frac{25}{84},\frac{25}{84})$& a,c \tabularnewline 
  0.857146 & 0 & $(0,0)+(\frac37,\frac37)$& a,c \tabularnewline   
\end{tabular}
\end{ruledtabular}
\caption{\label{table:fsdata75p}
  Similar as Table~\ref{table:fsdata53p} but for the $(7,5,+1)$ model.  Again, the corresponding root patterns $\Delta_\gamma$ are obscured by finite size effects.}
\end{table}
As in the $(7,3,+1)$ model the finite size data are consistent with the conjecture that the critical theory is a product of two sectors described by the $\mathcal{M}_{(5,6)}$ minimal model  with central charge $c_1=4/5$ and an $c=1$ rational CFT, the latter being a $\mathbf{Z}_2$ orbifold of a Gaussian model with radius $2R^2=21$.  The levels identified with solutions of the Bethe equations in Table~(\ref{table:fsdata75p}) contains all but one of the lowest ones predicted based on this conjecture.  By numerical diagonalization for systems of length up to $L=10$ we find that there is indeed another level in the finite size spectrum of both the clock and the anyon model which is consistent with the missing scaling dimension $X_2=\frac{2}{15}+\frac{8}{21}\simeq 0.5142857\ldots$:

\begin{center}
\begin{ruledtabular}
\begin{tabular}{c|ccccc|c|c}
    $L$ & $2$ & $4$ & $6$ & $8$ & $10$ & extrap. & conj. \\
    \hline
    & 0.68355511 
    & 0.58292862 
    & 0.56046603 
    & 0.54980798 
    & 0.54346327 
    & 0.52(1) 
    & $(\frac{1}{15},\frac{1}{15}) + (\frac{4}{21},\frac{4}{21})$
\end{tabular}
\end{ruledtabular}
\end{center}

Similarly as for the $(n,\ell_{FZ},J=(5,3,\pm1)$ models the numerical data characterizing the ground state phase diagram of the generic Hamiltonian $\mathcal{H}_{(7)}$ (\ref{hamil7_sphere}), see Fig.~\ref{fig:phasp_n7}, indicate that the integrable points $(7,3,+1)$ and $(7,5,+1)$ are located in an extended critical phase with effective central charge $c=\frac45+1$ and a non-universal dependence of the corresponding Fermi velocities on the coupling constants.


\paragraph*{\underline{$(n,\ell_{FZ},J)=(7,5,-1)$}.}
From the root density approach we obtain the ground state energy density for this model to be $\epsilon_\infty = -3.96535709066090$.  There are two branches of low lying excitations over the ground state with different Fermi velocities $v^{(F)}_1=7$, $v^{(F)}_2=7/5$.  Based on our extrapolation of the numerical finite size data for $C(L) \equiv -(6 L/\pi) (E_0(L)-L\epsilon_\infty)$ we conjecture the critical theory to be a product of two sectors with $c_1=1/2$ and $c_2=1$, respectively, giving $C(L) = v^{(F)}_1 c_1 + v^{(F)}_2 c_2 \to 49/10$, see Table~\ref{table:fsdata75m}.
\begin{table}[t]
\begin{ruledtabular}
\begin{tabular}{c|ccccc|c|c|c}
    $L$ 
        & $5$ & $6$ & $7$ & $8$ & $9$ & extrap. & conj. \\
    \hline
    $C(L)$     
               & $4.9297964$
               & $4.9211744$
               & $4.9157441$
               & $4.9121370$
               & $4.9096312$
    & $4.899(1)$ & $\frac12+1$ & a,c
  \\\hline
  $X(L)$ 
         &  0.1726247 
         &  0.1733539 
         &  0.1737921 
         &  0.1740760 
         &  0.1742703 
    & 0.1750(1) &  $(0,0)+(\frac{1}{16},\frac{1}{16})$ & a
  \\
         &  0.1940875 
         &  0.1959664 
         &  0.1970656 
         &  0.1977667 
         &  0.1982423 
    & 0.2000(3) & $(0,0)+(\frac{1}{14},\frac{1}{14})$ & a,c
  \\   
           &  0.8100552 
           &  0.8068032 
           &  0.8049255 
           &  0.8037371 
           &  0.8029350 
     & 0.8005(6) & $(0,0)+(\frac27,\frac27)$ & a,c
  \\   
          &  0.9284772 
          &  0.9274070 
          &  0.9267650 
          &  0.9263496 
          &  0.9260655 
     & 0.9250(1) & $(\frac{1}{16},\frac{1}{16})+(\frac{1}{56},\frac{1}{56})$ & a,c
  \\   
           &   1.0486155 
           &   1.0491055 
           &   1.0493694 
           &   1.0495292 
           &   1.0496342 
    & 1.050(2) & $(\frac{1}{16},\frac{1}{16})+(\frac{1}{16},\frac{1}{16})$ & a
  \\   
           &   1.0611437 
           &   1.0576541 
           &   1.0555890 
           &   1.0542630 
           &   1.0533600 
    & 1.052(5) & $(\frac{1}{16},\frac{1}{16})+(\frac{1}{16},\frac{1}{16})$ & a
  \\ 
           &   1.3284775 
           &   1.3274085 
           &   1.3267667 
           &   1.3263512 
           &   1.3260669 
    & 1.3250(1) & $(\frac{1}{16},\frac{1}{16})+(\frac{9}{56},\frac{9}{56})$ & a,c
  \\ 
           &   1.6118182 
           &   1.6001571 
           &   1.5933096 
           &   1.5889348 
           &   1.5859659 
    & 1.5751(2) & $(0,0)+(\frac{9}{16},\frac{9}{16})$ & a
  \\ 
           &   1.9180760 
           &   1.8811101 
           &   1.8591850 
           &   1.8451080 
           &   1.8355284 
    & 1.7999(1) & $(0,0)+(\frac{9}{14},\frac{9}{14})$ & a,c
\end{tabular}
\end{ruledtabular}
\caption{\label{table:fsdata75m}
  Similar as Table~\ref{table:fsdata53m} but for the $(7,5,-1)$ model.  The last column displays the conjectured contributions to the central charge and conformal weights of the two factors (Ising and $\mathbf{Z}_2$-orbifolded boson with compactification radius $2R^2=14$) to the finite size amplitudes $Y = v^{(F)}_1Y_1 + v^{(F)}_2Y_2$ (for $Y=C$, $X$), see Eqs.~(\ref{C2-extra}) and (\ref{X2-extra}).} 
\end{table}
Similarly, the data $X(L) = (L/2\pi)\left(E_n(L) - L\epsilon_\infty\right) + C(\infty)/12$ for the lowest excitations are consistent with those observed for the $(7,3,-1)$ model discussed above.  Therefore, we conjecture that the  $c=1$ factor is the a $\mathbf{Z}_2$-orbifold of a boson with compactification radius $2R^2=14$.
Together with our numerical data characterizing the ground state phase diagram of the generic $\mathcal{H}_{(7)}$ (\ref{hamil7_sphere}), see Fig.~\ref{fig:phasp_n7}, this indicates that the integrable points $(7,3,-1)$ and $(7,5,-1)$ are located in an extended critical phase with effective central charge $c=\frac12+1$.


\begin{acknowledgments}
This work has been carried out within the research unit \emph{Correlations in Integrable Quantum Many-Body Systems} (FOR2316).   Financial support by the Deutsche Forschungsgemeinschaft through grant no.\ Fr\,737/8-1 is gratefully acknowledged.
The numerical computations for this work were partially performed on the cluster system at Leibniz Universit\"at Hannover, Germany.
\end{acknowledgments}

\newpage
\begin{appendix}

\section{One-dimensional anyon chains from braided fusion categories}
\label{app:fuscat}
Algebraically anyonic theories can be described by braided tensor categories \cite{Kita06}.
A braided tensor category consists of a collection of objects
$\{\psi_{i}\}_{i\in\mathcal{I}}$ (including an identity) equipped with a
tensor product (fusion rules),
\begin{align*}
  \psi_{a} \tp \psi_{b} & \cong \bigoplus_{c} N_{ab}^{c} \psi_{c}
\end{align*}
where $N_{ab}^{c}$ are natural numbers (including zero).  In the special case
where $N_{ab}^{c}\in\{0,1\}$ the tensor category is said to be multiplicity free, a property which will be assumed for the remainder of the paper. Fusion can be represented graphically
\begin{align*}
\begin{tikzpicture}[scale=1.0]
	\tikzstyle{every node}=[minimum size=0pt,inner sep=0pt,style={sloped,allow upside down}]
	\tikzstyle{every loop}=[]
	\node (na) at (0.0,1.0) {$\psi_a$};
	\node (nb) at (1.0,1.0) {$\psi_b$};
	\node (nm) at (0.5,0.5) {};
	\node (nc) at (0.5,0.0) {$\psi_c$};
	\foreach \from/\to in {na/nm,nb/nm,nm/nc} \draw[middlearrow={latex}] (\from) -- (\to);
\end{tikzpicture}
\end{align*}
provided that $\psi_c$ appears in the fusion of $\psi_a$ and $\psi_b$.

We require associativity in our fusion i.e.\
\begin{align*}
  (\psi_{a} \tp \psi_{b}) \tp \psi_{c} & \cong \psi_{a} \tp (\psi_{b} \tp
  \psi_{c}), 
\end{align*}
which is governed by $F$-moves, also referred to as generalised 6-j symbols,
\begin{align*}
\begin{tikzpicture}[scale=1.0]
	\tikzstyle{every node}=[minimum size=0pt,inner sep=0pt,style={sloped,allow upside down}]
	\tikzstyle{every loop}=[]
	\node (ns) at (0.2,0.0) {$\psi_a$};
	\node (ne) at (2.8,0.0) {$\psi_e$};
	\node (n1t) at (1.0,0.9) {$\psi_b$};
	\node (n2t) at (2.0,0.9) {$\psi_c$};
	\node (n1b) at (1.0,0.0) {};
	\node (n2b) at (2.0,0.0) {};
	\node (l1) at (1.6,-0.3) {$\psi_d$};
	\foreach \from/\to in {ns/n1b,n1b/n2b,n2b/ne} \draw[middlearrow={latex}] (\from) -- (\to);
	\foreach \from/\to in {n1t/n1b,n2t/n2b} \draw[middlearrow={latex}] (\from) -- (\to);
	\node (l2) at (4.8,0.0) {$=\sum_{d'} (F^{abc}_{e})^{d}_{d'}$}; 
	\node (vs) at (6.9,-0.2) {$\psi_a$};
	\node (ve) at (9.5,-0.2) {$\psi_e$};
	\node (v1) at (7.7,0.9) {$\psi_b$};
	\node (v2) at (8.7,0.9) {$\psi_c$};
	\node (v12) at (8.2,0.3) {};
	\node (vb) at (8.2,-0.2) {};
	\node (l3) at (8.55,0.05) {$\psi_{d'}$};
	\foreach \from/\to in {vs/vb,vb/ve} \draw[middlearrow={latex}] (\from) -- (\to);
	\foreach \from/\to in {v1/v12,v2/v12,v12/vb} \draw[middlearrow={latex}] (\from) -- (\to);
\end{tikzpicture}
\end{align*}
For more than three objects different decompositions of the fusion can be
related by distinct series of $F$-moves. Their consistency for an arbitrary
number of anyons is guaranteed by the Pentagon equation satisfied by the
$F$-moves.
There also must be a mapping that braids two objects, $R: \psi_{a} \tp
\psi_{b} \rightarrow \psi_{b} \tp \psi_{a}$.  Here, however, we are not
concerned with such maps.

Given a consistent set of rules for the fusion we can construct a periodic
one-dimensional chain of $2L$ interacting `anyons' with topological charge
$\psi_j$.\footnote{Here we are only considering chains of even length, in
  general one may also consider chains of odd length.}  A basis vector for
such a model is defined graphically as
\begin{align*}
  \begin{tikzpicture}[scale=1.0]
  \tikzstyle{every node}=[minimum size=0pt,inner sep=0pt,style={sloped,allow upside down}]
	\tikzstyle{every loop}=[]
	\node (ns) at (0.2,0.0) {$\dots$}; 
	\node (ne) at (6.8,0.0) {$\dots$};
	\node (nebl) at (6.8,-0.3) {$\psi_{a_{2L}}$};
	\node (n1t) at (1.0,0.9) {$\psi_j$};
	\node (n2t) at (2.0,0.9) {$\psi_j$};
	\node (n3t) at (3.0,0.9) {$\psi_j$};
	\node (n5t) at (5.0,0.9) {$\psi_j$};
	\node (n6t) at (6.0,0.9) {$\psi_j$};
	\node (n1b) at (1.0,0.0) {};
	\node (n2b) at (2.0,0.0) {};
	\node (n3b) at (3.0,0.0) {};
	\node (n5b) at (5.0,0.0) {};
	\node (n6b) at (6.0,0.0) {};
	\node (nm1) at (3.4,0.0) {};
	\node (nm2) at (3.8,0.0) {};
	\node (nm3) at (4.2,0.0) {};
	\node (nm4) at (4.6,0.0) {};
	\node (n1bl) at (1.6,-0.3) {$\psi_{a_{1}}$};
	\node (n2bl) at (2.6,-0.3) {$\psi_{a_{2}}$};
	\node (n5bl) at (5.6,-0.3) {$\psi_{a_{2L-1}}$};
	\foreach \from/\to in {ns/n1b,n1b/n2b,n2b/n3b,n3b/nm1,nm2/nm3,nm4/n5b,n5b/n6b,n6b/ne} \draw[middlearrow={latex}] (\from) -- (\to);
	\foreach \from/\to in {n1t/n1b,n2t/n2b,n3t/n3b,n5t/n5b,n6t/n6b} \draw[middlearrow={latex}] (\from) -- (\to);
  \node (ns) at (9.0,0.3) {$\equiv \ket{a_{1},a_{2},...,a_{2L-1},a_{2L}}.$};
  \end{tikzpicture}
\end{align*}
As we are dealing with periodic models we always identify $a_{i}$ with
$a_{i+2L}$. Mathematically, we define the Hilbert space of the chain of $2L$
the anyons with charge $\psi_j$ to be the vector space spanned by
\begin{equation}
  \label{anybasis}
  \mathcal{B}_L^{(j)} = \left\{\ket{a_{1},a_{2},...,a_{2L}} |
       a_{i}\in\mathcal{I}\,\, \mbox{s.t.}\,\, N_{a_{i-1}j}^{a_{i}}=1  
     \right\}\,.
\end{equation}
This Hilbert space can be further decomposed into topological sectors based on the eigenvalues of a family of charges $\{Y_b\}_{b\in\mathcal{I}}$: these charges can be
measured by inserting an additional anyon of type $\psi_b$ into the system
which is then moved around the chain using the $F$-moves and finally removed
again \cite{FTLT07}.  The corresponding topological operator $Y_b$ has matrix
elements
\begin{equation}
  \label{Ytopo}
  \bra{a_{1}',a_{2}',...,a_{2L}'} Y_{b} \ket{a_{1},a_{2},...,a_{2L}} 
  = \prod_{i=1}^{2L} \left(F^{b
      a_{i}'j}_{a_{i+1}}\right)^{a_{i}}_{a_{i+1}'}\,,
  \quad b\in\mathcal{I}\,.
\end{equation}
The spectrum of these operators is known: their eigenvalues are given in terms
of the modular $\mathcal{S}$-matrix which diagonalizes the fusion rules
\cite{Kita06}.

The only other operators that we consider on this space are two-site projection
operators:
\begin{equation} 
  \label{eqnProjOp}
  p^{(b)}_{i}  =  \sum_{\ket{\bf{a}},\ket{\bf{a}'}\in\mathcal{B}_L^{(j)}}
  \left[\prod_{k\neq i} \delta_{a_{k}}^{a_{k}'}\right]
  \left(\bar{F}^{a_{i-1}jj}_{a_{i+1}}\right)^{b}_{a_{i}'}
  \left(F^{a_{i-1}jj}_{a_{i+1}}\right)^{a_{i}}_{b}
  \ket{\bf{a}'}\bra{\bf{a}}\,. 
\end{equation}
where $\bar{F}$ is the inverse $F$-move.
Note that the matrix elements of these operators depend on triples of
neighbouring labels $a_{i-1}a_{i}a_{i+1}$ in the fusion path but only the
middle one may change under the action of the $p^{(b)}_{i}$.
In terms of the local projection operators the global Hamiltonian describing
nearest-neighbour interactions between $\psi_j$ anyons is given by
\begin{equation}
  \label{anyhamil}
  \mathcal{H}({\boldsymbol\alpha})  =
  \sum_{i=1}^{2L}\left[\sum_{b\in\mathcal{I}} \alpha_{b}\,
    p_{i}^{(b)}\right].
\end{equation}
This generic description has obvious redundancies, for example, it is clear
the sum need not be over all $b\in\mathcal{I}$ but can instead by restricted
to $b$ such that $\psi_{j}\tp\psi_{j}\cong\psi_{b}\oplus\cdots$.  By construction the Hamiltonian commutes with the topological charges (\ref{Ytopo}).

\subsection*{Symmetries from monoidal equivalences}
Suppose that there exist two sets of $F$-moves, $F$ and $\widetilde{F}$,
together with the corresponding projection operators (\ref{eqnProjOp}).  $F$
and $\widetilde{F}$ are monoidally equivalent if they obey the relation
\begin{align*}
  \left(F^{abc}_{d}\right)^{e}_{f} & =
  \frac{u^{af}_{d}u^{bc}_{f}}{u^{ab}_{e}u^{ec}_{d}}
  \left(\widetilde{F}^{\nu(a)\nu(b)\nu(c)}_{\nu(d)}\right)^{\nu(e)}_{\nu(f)} 
 \end{align*}
where $u_{ab}^{c}\in\C$ and $\nu:\mathcal{I}\rightarrow\mathcal{I}$ is an
automorphism of the fusion rules,
\begin{align*}
  N_{ab}^{c} & = N_{\nu(a)\nu(b)}^{\nu(c)}.
\end{align*}
In the special case where the automorphism is trivial, i.e. $\nu=\mbox{id}$,
then $F$ and $\widetilde{F}$ are said to be gauge equivalent.
In this paper we say that $\widetilde{F}$ is monoidally related to $F$ via
$\nu$ (or, equivalently, $F$ is monoidally related to $\widetilde{F}$ via
$\nu^{-1}$).  This distinction is convenient as it allows us keep track of
permutations which arise.

It is natural to ask if one can relate models constructed from the different
$F$-moves.  Consider the invertible operator which maps states from the
Hilbert space of anyons with topological charge $\psi_j$ to that of anyons
with topological charge $\psi_{\nu(j)}$,
\begin{align}
  \bra{\bf{a}'} U \ket{\bf{a}} & = \left[\prod_{i=1}^{L}
    \frac{\delta_{\nu(a_{i})}^{a_{i}'}}{u^{a_{i}j}_{a_{i+1}}} \right] &
  \forall \ket{\bf{a}}\in\mathcal{B}_L^{(j)}\,,\,\,
  \ket{\bf{a}'}\in\mathcal{B}_L^{(\nu(j))} 
  \label{eqn:Udef}
\end{align}
This map provides an equivalence between different Hamiltonians,
\begin{align*}
  \mathcal{H}(\nu({\boldsymbol\alpha})) 
  & = U^{-1} \widetilde{\mathcal{H}}({\boldsymbol\alpha}) U, 
  & \mbox{where} \quad \nu(\alpha)_{a} = \alpha_{\nu(a)},
\end{align*}
where $\mathcal{H}({\boldsymbol\alpha})$ is the Hamiltonian built from $F$
acting on a chain of $\psi_{j}$-anyons, while
$\widetilde{\mathcal{H}}({\boldsymbol\alpha})$ is the Hamiltonian built from
$\widetilde{F}$ acting on a chain of $\psi_{\nu(j)}$-anyons.

If one considers a chain of $\psi_{j}$-anyons and an automorphism satisfying
$\nu(j)=j$ (for that $j$) then one can equate different models via the above
transformation. If it is also the case that $F=\widetilde{F}$, which amongst
other things necessitates $\mathcal{H}({\boldsymbol\alpha})=
\widetilde{\mathcal{H}}({\boldsymbol\alpha})$, then the monodial equivalence
implies that the parameter space of the model possesses a symmetry governed by
$\nu$. For instance, if $\nu$ is of order $n$, i.e. the minimum $n$ such that
$\nu^{n}=\mbox{id}$, then the model must contain a global $\mathbf{Z}_{n}$ symmetry.

\section{Thermodynamic limit of the integrable chains}
\label{app:thermo}
To analyze the properties of the integrable $\mathbf{Z}_n$ clock models and
$so(n)_2$ anyon chains in the thermodynamic limit, $L\to\infty$, the root
configurations solving the Bethe Eqs.~(\ref{baeFZ}) and (\ref{baeSO})
corresponding to the ground state and low lying excitations have to be
identified.
For small system sizes the Bethe roots can be obtained by direct
diagonalization of the transfer matrices: in the $so(n)_2$ case they can be
related to zeroes of the eigenvalues while in the $\mathbf{Z}_n$ case a
general non-uniform $R$-matrix has to be considered, see Ref.~\cite{BaSt90}.
Although this approach is limited by the available computational resources we
find that -- for sufficiently large $L$ -- Bethe roots can be grouped into
several characteristic patterns: a group of Bethe roots $\{u_j\}$ with
identical real part is said to form a pattern of type $\gamma$ if
\begin{align*}
  u_j \simeq u^{(\gamma)} + \mu_j + \delta\,,\qquad
    u^{(\gamma)}\in\mathbb{R}\,,\,\, \mu_j\in S_\gamma\,.
\end{align*}
For the Bethe equations (\ref{baeFZ}) and (\ref{baeSO}) we have to consider
the root patterns\footnote{This generalizes Albertini's conjecture for the
  $\ell_{FZ}=1$ clock models \cite{Albe94}.}
\begin{enumerate}
\item $(k,+)$-strings: $S_{(k,+)} =
  \left\{\left.\frac{i\pi}{2}\left(1-\frac{\ell_{FZ}}{n}\right)
      \left(t-\frac{k+1}{2}\right) \right| t=1\ldots k\right\}$\,,
\item $(k,-)$-strings: $S_{(k,-)} =
  \left\{\left.\frac{i\pi}{2}+\frac{i\pi}{2}\left(1-\frac{\ell_{FZ}}{n}\right)
      \left(t-\frac{k+1}{2}\right) \right| t=1\ldots k\right\}$\,,
\item $k$-multiplets: $S_{(k,m)} =
  \left\{\left. \pm\frac{i\pi}{4}+\frac{i\pi}{2}\frac{\ell_{FZ}}{n}
      \left(t-\frac{k+1}{2}\right) \right| t=1\ldots k\right\}$\,.
\end{enumerate}

Within the root density formalism \cite{YaYa69} the thermodynamic properties
of the models can be studied based on this classification of Bethe roots.
The solution to the Bethe equations corresponding the ground state of the
$(n,\ell_{FZ},J)$-models of length $L$ can be decomposed into $d_\gamma$
patterns of type $\gamma\in\Gamma$ extending over the entire real axis with
\begin{align*}
  \sum_{k=1}^{n-1} k\left( d_{(k,+)} + d_{(k,-)} + 2d_{(k,m)}\right) = (n-1)\,L\,.
\end{align*}
Their densities $\rho_\gamma(u)$ satisfy a system of coupled linear integral
equations
\begin{align}
  \label{igl-rho}
  D_{\gamma}\rho_{\gamma}(u) &
  = \rho^{(0)}_{\gamma}(u) - \sum_{\gamma'\in\Gamma}
  \int_{-\infty}^{\infty} \mathrm{d}v\, K_{\gamma,\gamma'}(u-v) \rho_{\gamma'}(v)
\end{align}
where $D_{(k,\pm)}=-J$, $D_{(k,m)}=-2J$, and
\begin{align*}
  K_{\gamma,\gamma'}(u) & = \sum_{\mu\in
    S_{\gamma}}\sum_{\mu'\in S_{\gamma'}}
  a\left(u;\frac{1}{2}-\frac{\ell_{FZ}}{2n} + \frac{\mu'-\mu}{i\pi}\right)\,, \\ 
  \rho^{(0)}_{\gamma}(u) & = \sum_{\mu\in S_{\gamma}}
  a\left(u;\frac{\ell_{FZ}}{4n} + \frac{\mu}{i\pi}\right) \,,
\end{align*}
with
\begin{align*}
  a(u;t) = -\frac{1}{\pi} \, 
  \frac{\sin 2\pi t}{\cosh2u - \cos2\pi t}\,.
\end{align*}
It is straightforward to solve the integral equations (\ref{igl-rho}) by
Fourier transform.  
In terms of the density functions $\rho_\gamma(u)$ the leading term of
the ground state energy (\ref{specFZ}), (\ref{specSO}) is
\begin{align*}
  E_0 &= L \epsilon_\infty +o(L^0)\,, &
  \epsilon_\infty = 
  & =  JL\pi \left\{ \sum_{\gamma\in\Gamma} \sum_{\mu\in S_{\gamma}}
    \int_{-\infty}^{\infty} \mathrm{d}u\, \rho_{\gamma}(u)
    a\left(u;\frac{\ell_{FZ}}{4n}+\frac{\mu}{i\pi}\right) \right\}\,.
\end{align*}
Low lying excitations are described by root configurations with finite
deviations $\Delta_\gamma=d_\gamma-d_\gamma^{(0)}$ from the
ground state distributions.  The have a linear dispersion with Fermi
velocities
\begin{equation*}
  v_{\gamma}^{(F)} =  \left. \frac{J}{2}
    \frac{\rho_{\gamma}'(u)}{\rho_{\gamma}(u)} \right|_{u\to\infty}. 
\end{equation*}
Using our finite size data as well we have been able to identify the
ground state configurations listed in Tables~\ref{table:thermo357},
\ref{table:thermo9}, \ref{table:thermo11}.  These proposals have been
checked against the ground state energy densities computed using the
\textsc{evomps} algorithm.

We observe that with $p$ being the integer satisfying $p \ell_{FZ} =
J\mod n$ the ground state configuration always contains at least one
of the patterns $(p,\pm)$.  For the model $(n,\ell_{FZ},J)=(n,1,-1)$,
i.e.\ $p=n-1$ the root configuration corresponding to the ground state
consists of only $(p,(-1)^{n+1)/2})$ patterns \cite{Albe92}.  The
Fermi velocity in this model is $v^{(F)}=n$ and from the finite size
corrections to the ground state energy we find the central charge of
the low energy effective theory to be $c=2\frac{n-1}{n+2}$.

\begin{table}[ht] 
\begin{ruledtabular}
\begin{tabular}{cccccc}
  $(n,\ell_{FZ},J)$ & 
    \multicolumn{4}{c}{Bethe root configuration}  & central charge(s)
    \tabularnewline
    & $S_{\gamma}$ & $D_{\gamma}$ & $d_{\gamma}/L$ & $v^{(F)}$ & 
  \tabularnewline \hline \hline
  $(3, 1, +1)$ &
    $S_{(1,+)}$ & $-1$ & $3/2$ & $3/2$   & $1$ \tabularnewline
  & $S_{(1,-)}$ & $-1$ & $1/2$ & $3/2$  
  \tabularnewline \hline
  $(3, 1, -1)$ &
    $S_{(2,+)}$ & $+1$ & $1$ & $3$  & $4/5$
  \tabularnewline \hline\hline
  $(5, 1, +1)$ &
    $S_{(1,+)}$ & $-1$ & $5/2$ & $5/4$ & $1$ \tabularnewline
  & $S_{(1,-)}$ & $-1$ & $3/2$ & $5/4$ 
  \tabularnewline \hline
  $(5, 1, -1)$ &
    $S_{(4,-)}$ & $+1$ & $1$ & $5$  & $8/7$
  \tabularnewline \hline
  $(5, 3, +1)$ &
    $S_{(2,+)}$ & $-1$ & $3/2$ & $5/2$ & $3/2$ \tabularnewline
  & $S_{(2,-)}$ & $-1$ & $1/2$ & $5/2$ 
  \tabularnewline \hline
  $(5, 3, -1)$ &
    $S_{(3,+)}$ & $+1$ & $1$   & $5$   & $1/2$ \tabularnewline
  & $S_{(1,m)}$ & $+2$ & $1/2$ & $5/3$ & $1$ 
  \tabularnewline \hline\hline
  $(7, 1, +1)$ &
    $S_{(1,+)}$ & $-1$ & $7/2$ & $7/6$ & $1$ \tabularnewline
  & $S_{(1,-)}$ & $-1$ & $5/2$ & $7/6$  
  \tabularnewline \hline
  $(7, 1, -1)$ &
    $S_{(6,+)}$ & $+1$ & $1$ & $7$ & $4/3$
  \tabularnewline \hline
  $(7, 3, +1)$ &
    $S_{(5,-)}$ & $-1$ & $1$ & $7$ &    $4/5$  \tabularnewline
  & $S_{(1,+)}$ & $-1$ & $1/2$ & $7/4$ & $1$ \tabularnewline
  & $S_{(1,-)}$ & $-1$ & $1/2$ & $7/4$  
  \tabularnewline \hline
  $(7, 3, -1)$ &
    $S_{(2,+)}$ & $+1$ & $2$ & $7/3$ & $3/2$ \tabularnewline
  & $S_{(2,-)}$ & $+1$ & $1$ & $7/3$  
  \tabularnewline \hline
  $(7, 5, +1)$ &
    $S_{(3,+)}$ & $-1$ & $3/2$ & $7/2$ & $9/5$ \tabularnewline
  & $S_{(3,-)}$ & $-1$ & $1/2$ & $7/2$  
  \tabularnewline \hline
  $(7, 5, -1)$ &
    $S_{(4,+)}$ & $+1$ & $1$ & $7$  & $1/2$ \tabularnewline
  & $S_{(1,m)}$ & $+2$ & $1$ & $7/5$ &  $1$
\end{tabular}
\end{ruledtabular}
\caption{\label{table:thermo357}
  Bethe root configurations for the ground states of the integrable models with $n=3$, $5$, and $7$.  Also shown are the Fermi velocities $v^{(F)}$ of the low-lying excitations and central charges of the  corresponding continuum theory.}  
\end{table}

\begin{table}[ht]
\begin{ruledtabular}
\begin{tabular}{ccccc}
  $(n,\ell_{FZ},J)$ & \multicolumn{4}{c}{Bethe root configuration} \\
    &$S_{\gamma}$ & $D_{\gamma}$ & $d_{\gamma}/L$ & $v^{(F)}$\\\hline
  $(9, 1, +1)$ &
    $S_{(1,+)}$ & $-1$ & $9/2$ & $9/8$  \tabularnewline
  & $S_{(1,-)}$ & $-1$ & $7/2$ & $9/8$  \\\hline
  $(9, 1, -1)$ &  $S_{(8,-)}$ & $+1$ & $1$ & $9$\\\hline
  $(9, 5, +1)$ & $S_{(2,+)}$ & $-1$ & $5/2$ & $9/4$  \tabularnewline
  &  $S_{(2,-)}$ & $-1$ & $3/2$ & $9/4$  \tabularnewline \hline
  $(9, 5, -1)$ & $S_{(7,-)}$ & $+1$ & $1$ & $9$  \tabularnewline
  &  $S_{(1,m)}$ & $+2$ & $1/2$ & $9/5$  \\\hline
  $(9, 7, +1)$ & $S_{(4,+)}$ & $-1$ & $3/2$ & $9/2$  \tabularnewline
  & $S_{(4,-)}$ & $-1$ & $1/2$ & $9/2$   \tabularnewline \hline
  $(9, 7, -1)$ & $S_{(5,+)}$ & $+1$ & $1$ & $9$  \tabularnewline
  &  $S_{(1,m)}$ & $+2$ & $3/2$ & $9/7$  
\end{tabular}
\end{ruledtabular}
\caption{\label{table:thermo9}Bethe root configurations for the ground
  states of some integrable models with $n=9$.  Also shown are
  the Fermi velocities $v^{(F)}$ of the low-lying excitations.}  
\end{table}

\begin{table}[ht]
\begin{ruledtabular}
\begin{tabular}{ccccc}
  $(n,\ell_{FZ},J)$ & \multicolumn{4}{c}{Bethe root configuration} \\
    &$S_{\gamma}$ & $D_{\gamma}$ & $d_{\gamma}/L$ & $v^{(F)}$
  \\\hline\hline
  $(11, 1, +1)$ & $S_{(1,+)}$ & $-1$ & $11/2$ & $11/10$  \tabularnewline
  &  $S_{(1,-)}$ & $-1$ & $9/2$ & $11/10$ \tabularnewline \hline
  $(11, 1, -1)$ &  $S_{(10,+)}$ & $+1$ & $1$ & $11$ \tabularnewline \hline
  $(11, 3, +1)$ &  $S_{(4,+)}$ & $-1$ & $1/2$ & $11/2$  \tabularnewline
  &  $S_{(4,-)}$ & $-1$ & $3/2$ & $11/2$  \tabularnewline
  &  $S_{(1,+)}$ & $-1$ & $1$ & $11/8$  \tabularnewline
  &  $S_{(1,-)}$ & $-1$ & $1$ & $11/8$  \tabularnewline \hline
  $(11, 5, +1)$ & $S_{(9,+)}$ & $-1$ & $1$ & $11$  \tabularnewline
  &  $S_{(1,+)}$ & $-1$ & $1/2$ & $11/6$  \tabularnewline
  &  $S_{(1,-)}$ & $-1$ & $1/2$ & $11/6$  \tabularnewline \hline
  $(11, 5, -1)$ &  $S_{(2,+)}$ & $+1$ & $3$ & $11/5$  \tabularnewline
  &  $S_{(2,-)}$ & $+1$ & $2$ & $11/5$ \tabularnewline \hline
  $(11, 7, -1)$ &  $S_{(3,+)}$ & $+1$ & $2$ & $11/3$  \tabularnewline
  &  $S_{(3,-)}$ & $+1$ & $1$ & $11/3$  \tabularnewline
  &  $S_{(1,m)}$ & $+2$ & $1/2$ & $11/7$  \tabularnewline \hline
  $(11, 9, +1)$ &  $S_{(5,+)}$ & $-1$ & $3/2$ & $11/2$  \tabularnewline
  &  $S_{(5,-)}$ & $-1$ & $1/2$ & $11/2$ \tabularnewline \hline
  $(11, 9, -1)$ &  $S_{(6,+)}$ & $+1$ & $1$ & $11$  \tabularnewline
  &  $S_{(1,m)}$ & $+2$ & $2$ & $11/9$  
\\\hline\hline
  $(13, 1, +1)$ & $S_{(1,+)}$ & $-1$ & $13/2$ & $13/12$  \tabularnewline
  &  $S_{(1,-)}$ & $-1$ & $11/2$ & $13/12$  \tabularnewline \hline
  $(13, 1, -1)$ &  $S_{(12,-)}$ & $+1$ & $1$ & $13$  \tabularnewline \hline
  $(13, 3, -1)$ & $S_{(4,+)}$ & $+1$ & $1$ & $13/3$  \tabularnewline
  &  $S_{(4,-)}$ & $+1$ & $2$ & $13/3$ \tabularnewline \hline
  $(13, 7, +1)$ &  $S_{(2,+)}$ & $-1$ & $7/2$ & $13/6$  \tabularnewline
  &  $S_{(2,-)}$ & $-1$ & $5/2$ & $13/6$  \tabularnewline \hline
  $(13, 9, +1)$ & $S_{(3,+)}$ & $-1$ & $5/2$ & $13/4$  \tabularnewline
  &  $S_{(3,-)}$ & $-1$ & $3/2$ & $13/4$  \tabularnewline \hline
  $(13, 11, +1)$ &  $S_{(6,+)}$ & $-1$ & $3/2$ & $13/2$  \tabularnewline
  &  $S_{(6,-)}$ & $-1$ & $1/2$ & $13/2$  
\end{tabular}
\end{ruledtabular}
\caption{\label{table:thermo11}Bethe root configurations for the ground
  states of some integrable models with $n=11$ and $13$.  Also shown are
  the Fermi velocities $v^{(F)}$ of the low-lying excitations.}  
\end{table}

\section{Rational CFTs with extended symmetries}
\label{app:rCFTs}
In the previous analysis \cite{FiFF14} of the integrable points of the $so(5)_2$ anyon chain $(n,\ell_{FZ},J)=(5,1,J)$ the continuum limit of the integrable chains has been found to be described by rational CFTs with extended chiral symmetry algebras (see \cite{BoSc95} and References therein) respecting the five-fold discrete ones in the anyon lattice model.
Based on this observation Casimir-type $\mathcal{W}$-algebras associated with the Lie-algebras $SO(n)=B_{(n-1)/2}$, $SO(2n)=D_n$, and the super Lie-algebra $OSp(1|n-1) = \mathcal{B}_{0,(n-1)/2}$
containing one fermionic generator with half-integer spin, are possible candidates for the low energy effective description of the models considered in this paper.  

Similarly, the scaling limit of the ferromagnetic, i.e.\ $J=-1$, FZ $n$-state clock models is known to be a $\mathbf{Z}_n$-invariant conformal field theory with parafermion currents \cite{ZaFa85,JiMO86}.  As rational CFTs, $\mathbf{Z}_n$ parafermions possess an extension of the Virasoro algebra to a $\mathcal{W}A_{n-1}$-algebra.

Below we list the central charges and conformal spectra of some field theories from the minimal series of these chiral symmetry algebras appearing in the low energy description of the $n$-state clock models and $so(n)_2$ anyon chains considered in this paper.

\subsection{Parafermions} 
The $\mathbf{Z}_n$ parafermion CFT has central charge $c_n = 2(k-1)/(k+2)$.  The conformal spectrum is known to be the set
\begin{align*}
  h_{\ell,m} &= \frac{1}{2}\frac{\ell(k-\ell)}{k(k+2)} +
  \frac{(\ell+m)(\ell-m)}{4k}\,, 
        \ \ \ \
        1\leq \ell\leq k\,,
        \ \ 
        -\ell\leq m\leq \ell\,,
        \ \ 
        \textrm{and}
        \ \
        \ell+m\equiv 0\ \mathrm{mod}\ 2\,,
\end{align*}
of conformal weights \cite{ZaFa85,GeQi87}.  Therefore, we get 
\begin{align}
  \label{cft_specZ3}
  k=3: \quad & c = \frac45\,,\quad
        h \in \left\{0,\frac{1}{15},\frac25,\frac23\right\}\,,\\
  \label{cft_specZ4}
  k=4: \quad & c = 1\,,\quad
        h \in \left\{0,\frac{1}{16},\frac1{12},\frac13,
           \frac9{16},\frac34,1\right\}\,,\\
  \label{cft_specZ5}
  k=5: \quad & c=\frac87\,,\quad
        h \in \left\{0,\frac{2}{35},\frac{3}{35},\frac27,
           \frac{17}{35},\frac{23}{35},\frac45,\frac67,\frac65 \right\}\,,\\  
  \label{cft_specZ7}
  k=7: \quad &c=\frac43\,,\quad
        h\in\left\{0, \frac{1}{21}, \frac{5}{63}, \frac{2}{21},
           \frac{2}{9}, \frac{8}{21}, \frac{11}{21},
           \frac{41}{63}, \frac23, \frac{16}{21}, \frac67,
           \frac{59}{63}, \frac{25}{21}, \frac43, \frac{10}{7}, 
           \frac{12}{7} \right\}\,.
\end{align}
As mentioned above, the $\mathbf{Z}_n$ parafermions appear in the minimal models of the $A$-series as $\mathcal{W}A_{n-1}(n+1,n+2)$.  Note that the operator content of the $\mathbf{Z}_3$ parafermion theory is a closed subset (under the fusion rules) of the scaling fields in the Virasoro minimal model $\mathcal{M}_{(5,6)}$.  The complete spectrum of this minimal model (the critical three-state Potts model) is
\begin{equation}
  \label{cft_specPotts3}
  h\in\left\{0,\frac{1}{40}, \frac{1}{15}, \frac18, \frac25,
       \frac{21}{40}, \frac23, \frac75, \frac{13}{8},3 \right\}\,.
\end{equation}
However, the presence of a representation with conformal weight $h=3$ and the existence of a non-diagonal partition function for this minimal model indicate that it's chiral symmetry algebra can be extended. In fact, considering $\mathcal{W}A_{2}(4,5)$ instead, which has a chiral symmetry algebra generated by the stress-energy tensor and a local chiral field of conformal weight $h=3$, we see that all representations of the minimal model with even Kac-labels drop out, so we are left with the pure $\mathbb{Z}_3$ parafermion spectrum.
$\mathbf{Z}_4$ parafermions also appear, due to $\hat{A}_3\cong\hat{D}_3$, in the $D$-series as $\mathcal{W}D_3(5,6)$ and are realized by the $\mathbf{Z}_2$-orbifold of a $U(1)$ boson with compactification radius $2R^2=3$ \cite{Ginsparg88,DVVV89}.

\subsection{Minimal models for $B_\ell=SO(2\ell+1)$}
The $\mathcal{W}B_2(5,7)$ CFT has central charge $c=\frac87$ and conformal weights \cite{FiFF14}
\begin{equation}
\label{cft_specWB2-57}
  h\in\left\{0,\frac{1}{28},{\frac {2}{35}},{\frac {3}{35}},{\frac {3}{28}},\frac14,\frac27,{
\frac {17}{35}},{\frac {15}{28}},{\frac {17}{28}},{\frac {23}{35}},\frac45
,\frac67,\frac65,{\frac {9}{7}},\frac74,{\frac {13}{7}},3\right\}\,.
\end{equation}
Note that the spectrum (\ref{cft_specZ5}) of the $\mathbf{Z}_5$ parafermion CFT is a subset of (\ref{cft_specWB2-57}).

The $\mathcal{W}B_3(7,9)$ CFT has central charge $c=\frac43$ and conformal weights \cite{FiFF14}
\begin{equation}
\label{cft_specWB3-79}
\begin{aligned}
  h\in&\left\{ 0,\frac{1}{24},\frac{1}{21},{\frac {5}{63}},\frac{2}{21},{\frac {7}{72}}, {\frac {5}{24}}, \frac29, \frac38,{\frac {8}{21}}, {\frac {11}{21}},{\frac {13}{24}},{\frac {43}{72}}, \right.\\
  & \quad\left.{\frac {41}{63}},\frac23,{\frac {17}{24}},{\frac {16}{21}}, \frac67,{\frac {59
}{63}},{\frac {25}{21}},{\frac {11}{9}}, \frac43,{\frac {10}{7}}, \frac53,{
\frac {12}{7}},{\frac {15}{8}},\frac73,3\right\}\,.
\end{aligned}
\end{equation}
Again, the spectrum (\ref{cft_specZ7}) of the $\mathbf{Z}_7$ parafermion CFT is contained in (\ref{cft_specWB3-79}).

\subsection{Minimal models for $D_\ell=SO(2\ell)$}
The $\mathcal{W}D_n(2n-1,2n)$ CFTs have central charge $c=1$. For $n=3$ the spectrum of conformal weights of this theory coincides with that of $\mathbf{Z}_4$ parafermions (\ref{cft_specZ4}) as discussed above.  The spectra for $n=5$, $7$, and some multiples thereof are
\begin{align}
\label{cft_specWD5-9-10}
  n=5: \quad &h\in\left\{0,\frac{1}{20},\frac{1}{16}, \frac15,{\frac {9}{20}},{\frac {9}{16}},\frac45,1,\frac54\right\}\,,\\
\label{cft_specWD7-1314}
  n=7: \quad &h\in\left\{0, \frac{1}{28}, \frac{1}{16}, \frac{1}{7},
           \frac{9}{28}, \frac{9}{16}, \frac{4}{7},
           \frac{25}{28}, 1, \frac97, \frac74 \right\}\,,\\
\label{cft_specWD10}
  n=10: \quad &h\in\left\{0, \frac{1}{40}, \frac{1}{16}, \frac{1}{10},
    {\frac {9}{40}},\frac25,{\frac {9}{16}},\frac58,{\frac {9}{
10}},1,{\frac {49}{40}},\frac85,{\frac {81}{40}},\frac52 \right\}\,,\\
\label{cft_specWD14}
  n=14: \quad &h\in\left\{0, {\frac {1}{56}},\frac{1}{16},\frac{1}{14},{\frac {9}{56}},\frac27,{\frac {25}{56}},{
\frac {9}{16}},{\frac {9}{14}},{\frac {7}{8}},1,{\frac {8}{7}},{\frac 
{81}{56}},{\frac {25}{14}},{\frac {121}{56}},{\frac {18}{7}},{\frac {
169}{56}},\frac72 \right\}\,,\\
\label{cft_specWD21}
  n=21:\quad & h\in\left\{0,{\frac {1}{84}},\frac{1}{21},\frac{1}{16},{\frac {3}{28}},{\frac {4}{21}},{\frac{25}{84}},\frac37,{\frac {9}{16}},{\frac {7}{12}},{\frac {16}{21}},{\frac{27}{28}},1,
  \right.\\
  &\nonumber\qquad\quad \left.
  {\frac {25}{21}},{\frac {121}{84}},{\frac {12}{7}},{\frac{169}{84}},\frac73,{\frac {75}{28}},{\frac {64}{21}},{\frac {289}{84}},{\frac{27}{7}},{\frac {361}{84}},{\frac {100}{21}},{\frac {21}{4}}
  \right\}\,.
\end{align}
We note that the spectra of these rational CFTs coincide with those of $\mathbf{Z}_2$-orbifolds of Gaussian models with compactification radii $2R^2=n$ indicating that the fields in the $\mathcal{W}D_n$ symmetry algebra are not independent. The orbifold models contain $n+7$ primary fields \cite{Ginsparg88,DVVV89}:
\begin{enumerate}
\item the identity with conformal weight $h_\mathbf{1} = 0$,
\item a marginal field $\Theta$ with conformal weight $h_\Theta= 1$,
\item two “degenerate” fields $\Phi^{1,2}$ with conformal weight $h_\Phi = \frac{n}{4}$,
\item the twist fields $\sigma_{1,2}$ and $\tau_{1,2}$, with conformal weights $h_\sigma = \frac{1}{16}$ and $h_\tau = \frac{9}{16}$, and
\item $(n-1)$ fields $\phi_k$, with $k = 1,2, . . . ,n-1$, with conformal weights $h_k =k^2/{4n}$.
\end{enumerate}
It is a well known fact that chiral symmetry algebras, which exist for generic values of the central charge, may collapse to smaller ones for certain values of the central charge. The best known example is the collapse of the $\mathcal{W}A_n$-algebra to the $\mathcal{W}A_2$-algebra at $c=1$. This phenomenon is called unifying $\mathcal{W}$-algebras \cite{BEHHH94}. A similar phenomenon happes with the $\mathcal{W}D_n(2n-1,2n)$-algebras at $c=1$, which all collapse down to mere $\mathcal{W}(2,4,n)$-algebras generated by the Virasoro stress-energy tensor and two further chiral primary fields of conformal dimensions $h=4$ and $h=n$, respectively. The reason is that, for special values of the central charge, several chiral primary fields become algebraically dependent due to the existence of additional null fields in the vacuum representation. 
Moreover, the $\mathcal{W}(2,4,n)$-algebras are in fact the maximally extended chiral symmetry algebras for $\mathbf{Z}_2$ orbifold Gaussian models with compactification radii $2R^2 = n$. It is not surprising that $\mathcal{W}$-algebras, which admit a rational CFT with finite representation content at central charge $c=1$, must shrink. The reason is that all rational CFTs at $c=1$ are known and their partition functions are classified \cite{Kiritsis88}.

\subsection{Minimal models for $\mathcal{B}_{0,\ell}=OSp(1|2\ell)$}
\label{app:rCFTs-WBf}
For non-simply-laced Lie-algebras alternative constructions of extended chiral symmetries are possible if one allows for generators with half-integer spin. In particular \cite{LuFa90}, one can construct an alternative $\mathcal{W}$-algebra for the $B_\ell$ series, i.e.\ for $SO(2\ell + 1)$, containing precisely one fermionic generator.  In general, these algebras can be realized from the Lie-superalgebras $\mathcal{B}_{0,\ell} = OSp(1|2\ell)$.  Therefore we denote them as $\mathcal{WB}_{0}$-algebras below. 
However, we note that these $\mathcal{W}$-algebras are not super-$\mathcal{W}$-algebras.

In the corresponding minimal series the $\mathcal{WB}_{0,n}(2n,2n+1)$ rational CFTs have central charge $c=1$.  The spectrum of conformal weights is
\begin{equation}
  \label{cft_specWBf2-45}
  \begin{aligned}
    h\in &\left\{0,\frac{1}{40},\frac{1}{16},\frac{1}{10},
      \frac{9}{40},\frac25,\frac{9}{16},\frac{5}{8},1\right\}\,
  \end{aligned}
\end{equation}
for $n=2$, and
\begin{equation}
  \label{cft_specWBf03-67}
  h \in \left\{0,\frac{1}{56},\frac{1}{16},\frac{1}{14},\frac{9}{56},\frac27,
          \frac{25}{56},\frac{9}{16},\frac{9}{14},\frac78, 1\right\}\,
\end{equation}
for $n=3$.
We note that these spectra are subsets of those for the $\mathcal{W}D_{10}(19,20)$ (\ref{cft_specWD10}) and $\mathcal{W}D_{14}(27,28)$ (\ref{cft_specWD14}), respectively.  

Similarly, the spectra of the $\mathcal{WB}_{0,n}(2n,2n+1)$ rational CFTs for general $n>2$ are contained in those of the $\mathcal{W}D_{4n+2}(8n+3,8n+4)$ models.  The representation with conformal weight $h=\frac{2n+1}{2}$ in the latter is part of the symmetry algebra in the $\mathcal{WB}_{0,n}$ case.  This may be an indication that the additional conformal weights in the $\mathcal{WD}_{4n+2}$ series are part of larger representations under the symmetry algebra of the $\mathcal{WB}_{0,n}$ series, as discussed in section IV.B with the example of the $\mathcal{WB}_{0,2}(4,5)$ model versus the $\mathcal{W}D_{10}(19,20)$ model. 

The $\mathcal{W}D_{4n+2}(8n+3,8n+4)$ models presumably admit non-diagonal partition functions in which characters of two representations, whose highest weights differ by a half-integer or an integer, are combined. In fact, the representations allowed in the $\mathcal{W}D_{4n+2}(8n+3,8n+4)$ models, but not contained in the spectra of the $\mathcal{WB}_{0,n}(2n,2n+1)$ rational CFTs, can all be seen to differ by a half-integer or an integer from one of the common ones.
This is easily seen in the above mentioned example, (\ref{cft_specWBf2-45}) and (\ref{cft_specWD10}). We find that the
spectra consist out of the orbifold weights
\begin{equation}
    h\in\left\{
        \frac{1}{16},\frac{9}{16},1\right\}
\end{equation}
common to all CFTs of these series, the weights
\begin{equation}
    h\in\left\{
        0,\frac{1}{40},\frac{1}{10},\frac{9}{40},\frac{2}{5},\frac{5}{8}
    \right\}
\end{equation}
common to both models, and finally the weights
\begin{equation}
    h\in\left\{
        \frac{9}{10}=\frac{2}{5}+\frac{1}{2},
        \frac{49}{40}=\frac{9}{40}+\frac{2}{2},
        \frac{8}{5}=\frac{1}{10}+\frac{3}{2},
        \frac{81}{40}=\frac{1}{40}+\frac{4}{2},
        \frac{5}{2}=0+\frac{5}{2}
    \right\}\,,
\end{equation}
which only appear in the $\mathcal{W}D_{10}(19,20)$ model, and differ by half-integers or integers from the common weights. The last weight, $h=\frac{5}{2}$, corresponds to a representation which is shifted by an half-integer above the vacuum representation $h=0$, and corresponds to a local chiral field which can be added to the chiral symmetry algebra. Indeed, the chiral symmmetry algebra of the $\mathcal{WB}_{0,2}(4,5)$ model does feature a generator of weight $h=\frac{5}{2}$.
Analogous results hold for the spectra of the pairs $\mathcal{W}D_{4n+2}(8n+3,8n+4)$ and
$\mathcal{WB}_{0,n}(2n,2n+1)$ for all $n$.  Among the representations in the common subset only the one with weight $h=\frac{2n+1}{8}$ does not have a corresponding representation in the shifted part of the  $\mathcal{W}D_{4n+2}(8n+3,8n+4)$ spectrum.

A further consequence of the existence of a generator of half-integer conformal weight is a two-fold degeneracy of all representations with $h\neq 0$. Let us explain this briefly: The  $\mathcal{WB}_{0,n}$ model has a symmetry algebra $\mathcal{W}(2,4,\ldots,\frac{2n+1}{2})$. Let us denote the generator of conformal weight 4 as $W$ and the generator of half-integer weight $\frac{2n+1}{2}$ as $Q$. The operator product expansion of the product $QQ$ will contain the field $W$. The corresponding quantum number $w$ for the zero mode $W_0$ of $W$ must satisfy a quadratic constraint to ensure associativity of the whole symmetry algebra. This leads to a relation $w=\pm h\,f(h)$ with an algebraic function $f$. Hence, for all $h\neq 0$, there are precisely two possible $w$-values. For the smallest case $n=2$, which we encountered in section IV.B, we can compute the function $f(h)$ explicitly, but this gets increasingly more difficult for larger $n$. We finally note that the associativity of the operator algebra for the $\mathcal{W}D_{4n+2}(8n+3,8n+4)$ does not yield obvious constraints of this type on the eigenvalues of the zero modes of the generators.

\end{appendix}

\newpage
%

\end{document}